\documentclass[a4paper,11pt,english,longbibliography]{article}
\pdfoutput=1

\usepackage{jheppub}
\usepackage{amsmath,amssymb,amsfonts,amsthm}
\usepackage{xcolor}
\usepackage{framed}
\definecolor{shadecolor}{gray}{0.95}
\usepackage{booktabs}
\usepackage{mathtools}
\definecolor{darkblue}{rgb}{0.1,0.1,.7}
\usepackage[]{amsmath}
\usepackage[]{graphicx}
\usepackage[]{latexsym}
\usepackage{amscd}
\usepackage[all,cmtip]{xy}
\usepackage{mathrsfs}
\usepackage{bbold}
\usepackage[margin=10pt,font=small,labelfont=bf]{caption}
\usepackage{simplewick}
\usepackage{changepage}
\usepackage[]{algorithm2e}
\usepackage{multirow}
\usepackage{slashed}
\usepackage[OT2,OT1]{fontenc}
\usepackage{braket}
\usepackage{tikz}
\usetikzlibrary{decorations.pathmorphing,circuits.logic.US,circuits.logic.IEC,fit,positioning,arrows,patterns,decorations.markings}
\tikzset{
	Witten diagram/.style={
		execute at begin picture={
			\draw[blue, line width=1.5pt] circle[radius=\pgfkeysvalueof{/tikz/Witten/radius}];
			\path node (X){\phantom{X}};
		},
		baseline={(X.base)}
	},
	vertex/.style={circle,fill,inner sep=1.5pt,node contents={}},
	Witten/.cd,
	radius/.initial=3cm
}

\theoremstyle{remark}

\makeatletter
\def\@fpheader{\ }
\makeatother

\title{RG flows in de Sitter: c-functions and sum rules}
\author{Manuel Loparco}
\affiliation{Fields and Strings Laboratory, Institute of Physics\\
École Polytechnique Fédéral de Lausanne (EPFL)\\
Route de la Sorge, CH-1015 Lausanne, Switzerland}
\emailAdd{manuel.loparco@epfl.ch}
\abstract{We study the renormalization group flow of unitary Quantum Field Theories on two-dimensional de Sitter (dS) spacetime. We prove the existence of two functions of the radius of dS that interpolate between the central charges of the UV and IR fixed points of the flow when tuning the radius $R$ while keeping the mass scales fixed. The first is constructed from certain components of the two-point function of the stress tensor evaluated at antipodal separation. The second is the spectral weight of the stress tensor in the $\Delta=2$ discrete series. This last fact implies that the stress tensor of any unitary QFT in dS$_2$ must interpolate between the vacuum and states in the $\Delta=2$ discrete series irrep. We verify that the c-functions are monotonic for intermediate radii in the free massive boson and free massive fermion theories, but we lack a general proof of said monotonicity. We derive a variety of sum rules that relate the central charges and the c-functions to integrals of the two-point function of the trace of the stress tensor and to integrals of its spectral densities. The positivity of these formulas implies $c^{\text{UV}}\geq c^{\text{IR}}$. In the infinite radius limit the sum rules reduce to the well known formulas in flat space. Throughout the paper, we prove some general properties of the spectral decomposition of the stress tensor in dS$_{d+1}$.}
\begin{document}
\maketitle

\newpage
\section{Introduction}
Unitary and Lorentz invariant quantum field theories (QFTs) in two dimensions describe renormalization group (RG) flows between two conformal field theories (CFT), one in the long distance (IR) regime, and one in the short distance (UV) regime. Zamolodchikov's seminal paper \cite{Zamolodchikov:1986gt} showed that to each flow one can assign a function which is monotonic in the scales of the theory and which asymptotes to the central charges of the two CFTs at the fixed points. The difference between the two central charges $\Delta c\equiv c^{\text{UV}}-c^{\text{IR}}$ is positive, a fact that is usually referred to as the c-theorem, and it can be related to sum rules involving integrals of observables computed along the flow \cite{Cardy:1988tj,Cappelli:1990yc,Karateev:2019ymz,Karateev:2020axc}
\begin{equation}
\begin{aligned}
    \Delta c=6\pi^2\int_{0}^{\infty}r^3dr\langle\Theta(r)\Theta(0)\rangle=12\pi\int_0^\infty \frac{ds}{s^2}\varrho_\Theta(s)\,,
    \label{eq:flatdeltac}
\end{aligned}
\end{equation}
where $r$ is a radial coordinate on the Euclidean plane, $\Theta$ is the trace of the stress tensor and $\varrho_\Theta$ is its spectral density over the $s=m^2>0$ unitary irreducible representations (UIRs) of the Poincar\'e group in two dimensions. 

The existence of a function that is monotonic under RG flows implies that the flows themselves are irreversible, giving a quantitative basis to the intuition that there is a loss of degrees of freedom when ``zooming out" and coarse graining in QFT. It is thus interesting to establish the existence of other RG-monotonic functions (also called c-functions) for QFTs in higher dimensions and on curved backgrounds, providing new general constraints on RG flows.

In \cite{Cardy:1988cwa}, Cardy conjectured that the one-point function of $\Theta$ integrated over a sphere could be a c-function in spacetimes with an even number of dimensions. This fact was proven in 4d by Komargodski and Schwimmer \cite{Komargodski:2011vj} and is called the $a$-theorem, since said integral isolates the coefficient of the Euler density in the trace anomaly of the UV and IR CFTs, usually denoted as $a$. In 3d, Casini and Huerta proved the $F$-theorem \cite{Casini:2012ei}, stating that the finite part of the free energy on a three-sphere satisfies $F^{\text{UV}}\geq F^{\text{IR}}$. This had been conjectured in \cite{Jafferis:2011zi}, and in \cite{Klebanov:2011gs,Giombi:2014xxa,Fei:2015oha} it was proposed that $\sin\left(\frac{\pi}{2}d\right)\log Z_{S^d}$ with $Z_{S^d}$ being the partition function of the theory on $S^d$, could be the generalization of $F$ to any dimension. While many checks and no counter examples are known, there is still no proof for this last statement. 

In this work, we focus on RG flows in reflection positive QFTs on a two-dimensional Euclidean sphere $S^2$ and unitary QFTs in two-dimensional de Sitter spacetime dS$_2$. The study of RG flows in dS has a long history, see for example \cite{Starobinsky:1986fx,Tsamis:1994ca,Tsamis:1996qm,Tsamis:1997za,Leblond:2002ns,Weinberg:2005vy,Weinberg:2006ac,Woodard:2008yt,Riotto:2008mv,Senatore:2009cf,Burgess:2009bs,Marolf:2010zp,Burgess:2010dd,Marolf:2011sh,Senatore:2012ya,Serreau:2013eoa,Serreau:2013psa,Anninos:2014lwa,LopezNacir:2016gzi,Gorbenko:2019rza,Baumgart:2019clc,Baumann:2019ghk,Green:2020txs,Heckelbacher:2022hbq,Chowdhury:2023arc,Chakraborty:2023eoq,Chakraborty:2023qbp,Cespedes:2023aal,Bzowski:2023nef,Agui-Salcedo:2023wlq,Cacciatori:2024zrv}. Leveraging recent advances in understanding non-perturbative unitarity \cite{DiPietro:2021sjt,Loparco:2023rug,Hogervorst:2021uvp,Penedones:2023uqc,Schaub:2023scu,DiPietro:2023inn} and analyticity \cite{Loparco:2023akg,Sleight_2020,Sleight_2021,Sleight_20202,Sleight_20212}, our main result is to prove the existence of two functions $c_1(R)$ and $c_2(R)$ which interpolate between the central charges at the fixed points of the RG flow as we tune the radius of $S^2/$dS$_2$ while keeping the mass scales of the theory fixed. At infinite radius we recover $c^{\text{IR}}$ and at vanishing radius $c^{\text{UV}}$. The two functions $c_1(R)$ and $c_2(R)$ are defined in terms of the correlation functions of the stress tensor. In the examples of the free massive boson and the free massive fermion, we find that these functions are also monotonic for intermediate $R$, although we do not have a theory-independent proof of this fact. We further consider the example of the massless Schwinger model, in which $c_1$ and $c_2$ match the same functions as in the free massive boson theory, hinting towards the fact that there exists a field redefinition that relates the two theories in dS, just as in flat space. 

\begin{figure}
\centering
\includegraphics[scale=1.8]{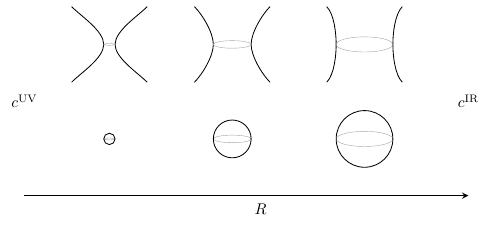}
\caption{Both in Euclidean and Lorentzian signature the functions $c_1$ and $c_2$ interpolate between the central charges of the CFTs at the endpoints of RG flows.}
\label{fig:spheres}
\end{figure}

The general point we advocate for in this work is that the radius of the sphere is a useful tunable parameter which can be used to follow RG flows in any QFT of interest.

\paragraph{Outline}
The paper is structured as follows. In Section \ref{sec:twodimensions}, we define $c_1(R)$ (\ref{eq:defcR}) and derive a sum rule (\ref{eq:sumrule2}) to compute $c^{\text{UV}}-c_1(R)$ in terms of an integral of the two-point function of the trace of the stress tensor over the chordal distance. We take a flat space limit and recover Cardy's sum rule \cite{Cardy:1988tj}, showing that $c_1(\infty)=c^{\text{IR}}$. Then, we use the K\"allén-Lehmann decomposition in de Sitter \cite{Bros:1995js,DiPietro:2021sjt,Hogervorst:2021uvp,Loparco:2023rug} to prove a sum rule for $c^{\text{UV}}-c_1(R)$ in terms of the spectral densities of the trace of the stress tensor (\ref{eq:sumrule3}). Its flat space limit reproduces the sum rule from \cite{Cappelli:1990yc}. Finally, we outline the technical assumptions under which $c_1(0)=c^{\text{UV}}$.

In Section \ref{sec:stresstensor}, we show that conservation greatly simplifies the spectral decomposition of the stress tensor (\ref{eq:stresstensorspectral}) in any number of dimensions. In two dimensions, we show that there are only three independent spectral densities: one for the principal series, one for the complementary series and one for the $\Delta=2$ discrete series (\ref{eq:stresstensorspectral2d}). The latter quantity is precisely $c_2(R)$. We show that $c_2(\infty)=c^{\text{IR}}$ and outline the assumption under which $c_2(0)=c^{\text{UV}}$, showing that it is more general than the analogous criterion for $c_1(R)$. We prove sum rules for $c^{\text{UV}}-c_2(R)$ in terms of integrals of the other spectral densities (\ref{eq:newrule}) and in terms of an integral of the two-point function of the trace of the stress tensor in position space (\ref{eq:newsumrule}). We also show sum rules which compute $c^{\text{UV}}$ (\ref{eq:cuvsum}) and $c_1(R)$ (\ref{eq:crsum}) independently.

In Section \ref{sec:examples}, we verify all our sum rules in the cases of a free massive boson and a free massive fermion. We find that $c_2(R)$ is monotonic and interpolates between the two central charges in both cases. The divergences associated with massless scalars in de Sitter make it so that the technical assumption at the basis of the proof that $c_1(0)=c^{\text{UV}}$ is violated. The theory of a massless fermion in dS, instead, is devoid of IR divergences and $c_1(R)$ still interpolates between $c^{\text{UV}}$ and $c^{\text{IR}}$ in the massive fermion flow. We comment on the fact that the massless Schwinger model has the same $c_1$ and $c_2$ functions as the free massive boson theory.

In Section \ref{sec:discussion}, we conclude and discuss some open questions.
\section{The first c-function and its sum rules}
\label{sec:twodimensions}
In this section we define $c_1(R)$ and we provide sum rules to compute $c^{\text{UV}}-c_1(R)$, checking that the flat space limit reproduces the known formulas from \cite{Cardy:1988tj,Cappelli:1990yc}. The techniques we use here closely parallel what was done in \cite{Meineri:2023mps} to derive sum rules for $c^{\text{UV}}$ in Anti de Sitter. The broad logic of this section is the following
\begin{itemize}
    \item \textbf{Assumptions}: the stress tensor is conserved and its two-point function reduces to the CFT form when the two points approach each other. In particular, its normalization constant is $c^{\text{UV}}$.
    \item \textbf{Construction}: we construct a differential equation (\ref{eq:diffeq}) involving the two-point function of the stress tensor. We choose an ansatz for the unknowns in (\ref{eq:diffeq}) in terms of components of the stress tensor two-point function and impose as a boundary condition that we retrieve $c^{\text{UV}}$ at coincident points.
    \item \textbf{Result}:  We find a solution to the differential equation and we integrate it. This lands us on a sum rule which returns $c^{\text{UV}}$ minus a quantity which we call $c_1(R)$. In the rest of the section we prove $c_1(R)$ interpolates between $c^{\text{UV}}$ and $c^{\text{IR}}$.
\end{itemize}
\subsection{Preliminaries}
\label{subsec:preliminaries}
We are going to treat both the Euclidean and Lorentzian cases together. The $S^{d+1}$ and dS$_{d+1}$ can be embedded in $\mathbb{R}^{1,d+1}$ as follows
\begin{equation}
	\pm(Y^0)^2+(Y^1)^2+\ldots+(Y^{d+1})^2=R^2\,.
 \label{eq:embedding}
\end{equation}
Throughout this and the following sections we will write ``de Sitter" to mean both the Lorentzian and Euclidean manifolds, to which all derivations apply.

We introduce the two-point invariant
\begin{equation}
	\sigma\equiv \frac{1}{R^2}Y_1\cdot Y_2
\end{equation}
where the dot is either $Y_1\cdot Y_2=\eta_{AB}Y_1^AY_2^B$ or $Y_1\cdot Y_2=\delta_{AB}Y_1^AY_2^B$ depending on the signature of choice, and the indices are $A=0,\ldots,d+1$. For now, we will set $R=1$ and then restore it when it is convenient. Operators can be lifted to embedding space and are related to their local counterparts in some coordinates $y^\mu$ with $\mu=0,1,\ldots d$ as follows
\begin{equation}
	T^{A_1\ldots A_J}=\frac{\partial Y^{A_1}}{\partial y^{\mu_1}}\cdots\frac{\partial Y^{A_J}}{\partial y^{\mu_J}}T^{\mu_1\ldots \mu_J}\,.
 \label{eq:embtoloc}
\end{equation}
The induced metric in embedding space and the covariant derivative are 
\begin{equation}
    G^{AB}=\eta^{AB}-Y^AY^B\,, \qquad 
    \nabla^A=\partial_Y^A-Y^A(Y\cdot\partial_Y)\,.
    \label{eq:metricderivs}
\end{equation}
The proof of the existence of $c_1(R)$ starts from considering the two-point function of the stress tensor on the Bunch-Davies vacuum, with the following choice of normalization\footnote{In our conventions the metric $g_{\mu\nu}$ is dimensionless, so $T_{\mu\nu}$ has mass dimensions $d+1$, as does the trace $\Theta$. Furthermore, we will consider the tensor structures (\ref{eq:tensorstructs}) dimensionless and the functions $T_i$ dimensionful.}
\begin{equation}
    T_{\mu\nu}\equiv-\frac{2}{\sqrt{|g|}}\frac{\delta S}{\delta g^{\mu\nu}}\,.
    \label{eq:defTmunu}
\end{equation}
Let us for now stay in general dimension $d+1$. By group theory, the two-point function of any spin 2 symmetric tensor can be decomposed into 5 linearly independent tensor structures
\begin{equation}
	\langle T^{AB}(Y_1)T^{CD}(Y_2)\rangle=\sum_{i=1}^5 \mathbb{T}_i^{ABCD}T_i(\sigma)\,.
 \label{eq:twopointstructs}
\end{equation}
The tensor structures we choose are, specifically,
\begin{equation}
\begin{aligned}
    \mathbb{T}_1^{ABCD}&=V_1^{A}V_1^{B}V_2^{C}V_2^{D}\,,\\
    \mathbb{T}_2^{ABCD}&=V_1^{A}V_1^{B}G_2^{CD}+G_1^{AB}V_2^{C}V_2^{D}\,,\\
    \mathbb{T}_3^{ABCD}&=-V_1^{A}V_2^{C}G_{12}^{BD}-V_1^{B}V_2^{C}G_{12}^{AD}-V_1^{A}V_2^{D}G_{12}^{BC}-V_1^{B}V_2^{D}G_{12}^{AC}\,,\\
     \mathbb{T}_4^{ABCD}&=G_1^{AB}G_2^{CD}\,,\\
      \mathbb{T}_5^{ABCD}&=G_{12}^{AD}G_{12}^{BC}+G_{12}^{AC}G_{12}^{BD}\,.
    \label{eq:tensorstructs}
\end{aligned}
\end{equation}
with
\begin{equation}
\begin{aligned}
    V_1^A&=Y_2^A-(Y_1\cdot Y_2)Y_1^A\,, \qquad &&V_2^A=Y_1^A-(Y_1\cdot Y_2)Y_2^A\,, \\
    G_1^{AB}&=\eta^{AB}-Y_1^AY_1^B\,,
    \qquad &&G_2^{AB}=\eta^{AB}-Y_2^AY_2^B\,,\\
    G_{12}^{AB}&=\eta^{AB}-\frac{Y_2^AY_1^B}{Y_1\cdot Y_2}\,.
    \label{eq:GVblobs}
\end{aligned}
\end{equation}
All of the tensors in (\ref{eq:GVblobs}) are transverse, so that we force the stress tensor to be tangential to the surface (\ref{eq:embedding}):
\begin{equation}
    V_i\cdot Y_i=G_i^{AB}Y_{i,A}=G_{12}^{AB}Y_{1,A}=G^{AB}_{12}Y_{2,B}=0\,.
\end{equation}
The connected two-point function of the trace $\Theta\equiv T^A_{\ A}$ will then be given by
\begin{equation}
\begin{aligned}
    \langle\Theta(Y_1)\Theta(Y_2)\rangle
    =&(\sigma^2-1)^2T_1(\sigma)+2(d+1)(1-\sigma^2)T_2(\sigma)+4\left(\frac{1}{\sigma}-\sigma\right)T_3(\sigma)\\
    &+(d+1)T_4(\sigma)+2\left(d+\frac{1}{\sigma^2}\right)T_5(\sigma)\,.
    \label{eq:2pttrace}
\end{aligned}
\end{equation}
The coincident point limit probes the CFT in the UV fixed point of the RG flow defined by our QFT. In particular, in that limit the two-point function (\ref{eq:twopointstructs}) has to reduce to the CFT two-point function of the stress tensor in the UV. This is uniquely fixed by symmetry and conservation up to a constant that is proportional to $c^{\text{UV}}$. In some Riemann normal coordinates $x^\mu$, this means
\begin{equation}
	\lim_{x\to0}\langle T^{\mu\nu}(x)T^{\varrho\sigma}(0)\rangle\approx
    \frac{c_T^{\text{UV}}}{x^{2d+2}}\left[\frac{1}{2}\left(w_{\mu\varrho}w_{\nu\sigma}+w_{\mu\sigma}w_{\nu\varrho}\right)-\frac{1}{d+1}\eta_{\mu\nu}\eta_{\varrho\sigma}\right]\,,
    \label{eq:flatCFT}
\end{equation}
with
\begin{equation}
    w_{\mu\nu}\equiv\eta_{\mu\nu}-2\frac{x_{\mu} x_{\nu}}{x^2}\,.
\end{equation}
where $c_T^{\text{UV}}$ is the normalization of the stress tensor two-point function in the UV, which in two dimensions is related to the central charge as follows
\begin{equation}
	c_T=\frac{1}{2\pi^2}c\,.
\end{equation}
This matching in the UV implies that the $T_i(\sigma)$ functions have the following behaviors at coincident points (see appendix (\ref{subsec:coincident}) for more details on how to derive this)
\begin{equation}
\begin{aligned}
    &T_1\approx\frac{4c_T^{\text{UV}}}{x^{2d+6}}\,, \qquad T_2\sim o(x^{-2d-2})\,,\qquad T_3\approx -\frac{c_T^{\text{UV}}}{x^{2d+4}}\,,\\
    &T_4\approx-\frac{c_T^{\text{UV}}}{d+1}\frac{1}{x^{2d+2}}\,,\qquad T_5\approx\frac{c_T^{\text{UV}}}{2}\frac{1}{x^{2d+2}}\,.
    \label{eq:coincidentTT}
\end{aligned}
\end{equation}
When defining the stress tensor through (\ref{eq:defTmunu}), we effectively impose it to be conserved at the fixed points, but we allow for the presence of local contact terms in its expectation values. To be more precise, (\ref{eq:coincidentTT}) should also include contact terms in the form of delta functions and their derivatives, such as is done explicitly in \cite{Hartman:2023ccw,Hartman:2023qdn}. All these terms would drop out of the sum rules we derive, and thus we do not report their explicit forms here.
\paragraph{Two dimensions} Effectively, in two dimensions (\ref{eq:twopointstructs}) is a redundant decomposition, since there are only 4 linearly independent tensor structures. This can be seen from the fact that necessarily
\begin{equation}
    W^{ABCD}_{\ \ \ \ EF}\equiv Y_1^{[A}Y_2^B\delta_E^C\delta_F^{D]}=0\,, \qquad \text{when}\ d+1=2\,.
\end{equation}
It is possible to check, then, that the equation $W^{A_1A_2CD}_{\ \ \ \ EF}W^{B_1B_2EF}_{\ \ \ \ CD}=0$ with $(A_1B_1)$ and $(A_2B_2)$ symmetrized, is equivalent to 
\begin{equation}
     -\frac{2}{\sigma^4}\mathbb{T}_1-\frac{2}{\sigma^2}\mathbb{T}_2+\frac{1}{\sigma^3}\mathbb{T}_3+2\frac{1-\sigma^2}{\sigma^2}\mathbb{T}_4-\frac{1-\sigma^2}{\sigma^2}\mathbb{T}_5=0\,,
\end{equation}
where we suppressed all the indices on the $\mathbb{T}_i$ to avoid clutter. This shows indeed that the tensor structures are degenerate in two dimensions. The $T_i(\sigma)$ functions are then defined up to a common shift by a generic function $g(\sigma)$
\begin{equation}
\begin{aligned}
    T_1(\sigma)&\sim T_1(\sigma)-\frac{2}{\sigma^4}g(\sigma)\,, \qquad &&T_2(\sigma)\sim T_2(\sigma)-\frac{2}{\sigma^2}g(\sigma)\\
    T_3(\sigma)&\sim T_3(\sigma)+\frac{1}{\sigma^3}g(\sigma)\,, \qquad &&T_4(\sigma)\sim T_4(\sigma)-2\frac{\sigma^2-1}{\sigma^2}g(\sigma)\,,\\
    T_5(\sigma)&\sim T_5(\sigma)+\frac{\sigma^2-1}{\sigma^2}g(\sigma)\,.
    \label{eq:redundancy}
\end{aligned}
\end{equation}
We thus construct four quantities which are invariant under this shift\footnote{The precise overall $\sigma$-dependent coefficient of each $\mathcal{T}_i$ function was chosen a posteriori after having derived their spectral decompositions (\ref{eq:TTTTspectrals}) in such a way that they would not diverge at antipodal separation $\sigma=-1$.}
\begin{equation}
    \begin{aligned}
	\mathcal{T}_1(\sigma)&\equiv(1-\sigma^2)\left[\frac{\sigma^2}{2}T_1(\sigma)-\frac{1}{2}T_2(\sigma)\right]\,,\qquad &&\mathcal{T}_3(\sigma)\equiv\frac{1}{2}T_4(\sigma)-(1-\sigma^2)\sigma T_3(\sigma)\\
	\mathcal{T}_2(\sigma)&\equiv(1-\sigma^2)\left[-\frac{1}{2}T_2(\sigma)-\sigma T_3(\sigma)\right]\,,\qquad &&\mathcal{T}_4(\sigma)\equiv\frac{1}{2}T_4(\sigma)+T_5(\sigma)\,.
        \label{eq:TtotildeT}
    \end{aligned}
\end{equation}
In this basis, the two-point function of the trace of the stress tensor has the following expression, in two dimensions
\begin{equation}
\begin{aligned}
G_\Theta(\sigma)=&\frac{2}{\sigma^2}\left((1-\sigma^2)\mathcal{T}_1(\sigma)-(1+3\sigma^2)\mathcal{T}_2(\sigma)+(3\sigma^2-1)\mathcal{T}_3(\sigma)+(1+\sigma^2)\mathcal{T}_4(\sigma)\right)
\end{aligned}
\end{equation}
where we introduced the notation
\begin{equation}
	G_\Theta(\sigma)\equiv\langle\Theta(Y_1)\Theta(Y_2)\rangle\,.
\end{equation}
Notice that the regularity of $G_\Theta(\sigma)$ at $\sigma=0$, which is not a special configuration on the sphere or in de Sitter, implies the following relation for the $\mathcal{T}_i$ functions
\begin{equation}
    \mathcal{T}_1(0)+\mathcal{T}_4(0)=\mathcal{T}_2(0)+\mathcal{T}_3(0)\,.
\end{equation}
\subsection{Proof of the position space sum rule}
\label{sec:proofposition}
For the purposes of our proof, we want to find a kernel $r(\sigma)$ and a function $C(\sigma)$ such that 
\begin{equation}
	r(\sigma)G_\Theta(\sigma)=\frac{d}{d\sigma}C(\sigma)\,.
 \label{eq:diffeq}
\end{equation}
Moreover, to extract the central charge, we would like to have $C(1)=c^{\text{UV}}$ up to contact terms, such that integrating both sides of (\ref{eq:diffeq}) will give us a sum rule. For this to work, necessarily $r(1)=0$ to kill the divergence of $G_\Theta(\sigma)$ at coincident points. 

In order to solve (\ref{eq:diffeq}), we use the following ansatz with four unknown functions $g_i(\sigma)$, purely motivated by the fact that it works\footnote{This choice was inspired by \cite{Meineri:2023mps} where a similar construction led to sum rules for $c^{\text{UV}}$ in AdS$_2$.}
\begin{equation}
\label{eq:Cansatz}
	C(\sigma)=\sum_{i=1}^4 g_i(\sigma)\mathcal{T}_i(\sigma)\,.
\end{equation}
Then, we impose the conservation of the stress tensor
\begin{equation}
	\nabla_A\langle T^{AB}(Y_1)T^{CD}(Y_2)\rangle=0\,.
\end{equation}
This induces three linearly independent scalar differential equations on the $T_i$ functions, which we obtain by multiplying with three linearly independent projectors, see (\ref{eq:conservationeqs}) for details. Using (\ref{eq:TtotildeT}), the conservation equations transform into three differential equations for the $\mathcal{T}_i$'s (\ref{eq:conservetildeT}).

Call $E_i=0$ with $i=1,2,3$ the three conservation equations (\ref{eq:conservetildeT}). Then, we introduce three unknown functions $q_i(\sigma)$ and say
\begin{equation}
	r(\sigma)G_\Theta(\sigma)-\frac{d}{d\sigma}C(\sigma)=\sum_{i=1}^3 q_i(\sigma)E_i\,.
\end{equation}
We impose that this equation be true for any $\sigma$, $\mathcal{T}_i(\sigma)$ and $\mathcal{T}'_i(\sigma)$. We treat $\mathcal{T}_i$ and $\mathcal{T}'_i$ as independent from each other, and imposing that their coefficients are zero gives 8 differential equations with 8 unknown functions, namely $r(\sigma)$, $g_1(\sigma),\ldots,g_4(\sigma)$ and $q_1(\sigma),q_2(\sigma),q_3(\sigma)$. We find three solutions which we report in (\ref{eq:threesols}). Only one of them has $C(1)=c^{\text{UV}}$ up to contact terms and $r(1)=0$. It has kernel
\begin{equation}
\begin{aligned}
    r(\sigma)&=8\pi^2\left[1-\sigma\left(\log\left(\frac{1+\sigma}{2}\right)+1\right)\right]\,,
\end{aligned}
\end{equation}
and associated function $C(\sigma)$ 
\begin{equation}
\begin{aligned}
	C(\sigma)=\frac{8\pi^2}{\sigma^2}\Big[&2(1-\sigma^2)^2\log(\zeta)\mathcal{T}_1(\sigma)\\
 &+2(\sigma^2(1-\sigma)^2+(\sigma^2-1)(2\sigma^2+1)\log(\zeta))\mathcal{T}_2(\sigma)\\
&+(\sigma(1-\sigma)^2(1-2\sigma)+2(2\sigma^2-1)(1-\sigma^2)\log(\zeta))\mathcal{T}_3(\sigma)\\
&-(\sigma(\sigma-1)^2+2(\sigma^2-1)\log(\zeta))\mathcal{T}_4(\sigma)\Big]\,,
\end{aligned}
\end{equation}
where $\zeta\equiv\frac{1+\sigma}{2}$. The fact that $\lim_{\sigma\to1}C(\sigma)=c^{\text{UV}}$ up to contact terms can be checked by using (\ref{eq:TtotildeT}) and (\ref{eq:coincidentTT}). Importantly, at antipodal separation, we have  
\begin{equation}
    C(-1)=32\pi^2R^4\left(T_5(-1)-T_4(-1)\right)=32\pi^2R^4\left(\mathcal{T}_4(-1)-3\mathcal{T}_3(-1)\right)\,,
\end{equation}
where we used the fact that the $T_i(\sigma)$ and $\mathcal{T}_i(\sigma)$ functions cannot diverge at $\sigma=-1$ and that $\mathcal{T}_2(-1)=0$, both facts which we prove in total generality in Appendix \ref{subsec:Tifuncs}, and we restored factors of the radius.
This is our $c_1(R)$, and we claim it interpolates between $c^{\text{UV}}$ and $c^{\text{IR}}$ as we change the radius of $S^2/$dS$_2$
\begin{equation}
    c_1(R)\equiv C(-1)\,.
    \label{eq:defcR}
\end{equation}
We will prove that the end-points of $c_1(R)$ are $c^{\text{UV}}$ and $c^{\text{IR}}$, and we will verify in examples in Section \ref{sec:examples} that $c_1(R)$ is a non-increasing function of the radius in between. 

Let us emphasize that, in a given QFT, each $\mathcal{T}_i$ function depends on the mass scales of the theory $\{m_k\}$ through dimensionless products such as $m_kR$ and $m_i/m_j$, hence the dependency on the radius of $c_1(R)$.

Integrating both sides of (\ref{eq:diffeq}) over the domain of the normalized inner product on the sphere $\sigma\in[-1,1)$, while being careful to avoid contact terms at $\sigma=1$, we get to one of our main results
\begin{shaded}
\begin{equation}
    c^{\text{UV}}-c_1(R)=8\pi^2\int_{-1}^{1} d\sigma\left[1-\sigma\left(\log\left(\frac{1+\sigma}{2}\right)+1\right)\right]R^4G_\Theta(\sigma)\,,
    \label{eq:sumrule2}
\end{equation}
\end{shaded}
\noindent where we restored the necessary factors of the radius. 

Let us note that, in this form, this sum rule is analogous to what was obtained in two-dimensional EAdS in \cite{Meineri:2023mps}\footnote{For a direct comparison, use $\sigma_{\text{here}}=-2\xi_{\text{there}}-1$.}
\begin{equation}
	 c^{\text{UV}}=8\pi^2\int_{-\infty}^{-1} d\sigma\left[-1-\sigma\left(\log\left(\frac{1-\sigma}{2}\right)+1\right)\right]R^4G_\Theta(\sigma)\,, \qquad \text{in AdS$_2$}\,.
\end{equation}
Notice the slightly different kernel, the different integration domain and the fact that the information about the intermediate flow is lost in the AdS case. 
\paragraph{Flat space limit}
Let us show that (\ref{eq:sumrule2}) reduces to (\ref{eq:flatdeltac}) in the flat space limit, thus proving that $c_1(\infty)=c^{\text{IR}}$. We start from the flat slicing coordinates $ds^2=R^2\frac{-\mathrm{d}\eta^2+\mathrm{d}y^2}{\eta^2}$ and we choose conventions in which the metric is dimensionless. The flat space limit is achieved by taking $\eta\to t-R$ and $y\to x$ and then taking $R\to\infty$. Then the metric becomes $ds^2=-\mathrm{d}t^2+\mathrm{d}x^2$ and
\begin{equation}
    \sigma=\frac{\eta_1^2+\eta_2^2-(y_1-y_2)^2}{2\eta_1\eta_2}\sim 1-\frac{-(t_1-t_2)^2+(x_1-x_2)^2}{2R^2}\equiv1-\frac{r^2}{2R^2}\,.
\end{equation}
Our formula (\ref{eq:sumrule2}) then changes as follows
\begin{equation}
\begin{aligned}
    c^{\text{UV}}-\lim_{R\to\infty}c_1(R)&=8\pi^2\lim_{R\to\infty}\int_{-1}^{1}d\sigma\ \left[1-\sigma\left(\log\left(\frac{1+\sigma}{2}\right)+1\right)\right]R^4G_\Theta(\sigma)\\
    &=6\pi^2\lim_{R\to\infty}\int_0^{2R}r^3dr\ G_\Theta\left(1-\frac{r^2}{2R^2}\right)\\
    &=6\pi^2\int_0^\infty r^3dr \langle\Theta(r)\Theta(0)\rangle^{\text{flat}}\,,
    \label{eq:flatspacelimitpos}
\end{aligned}
\end{equation}
The last form precisely matches (\ref{eq:flatdeltac}), implying that
\begin{equation}
	\lim_{R\to\infty}c_1(R)=c^{\text{IR}}\,.
\end{equation}
\subsection{A sum rule in terms of spectral densities}
\label{subsec:proofspectral}
We are interested in phrasing (\ref{eq:sumrule2}) in terms of an integral over the spectrum of the theory. To do that, we are going to use the fact that the two-point function of the trace of the stress tensor in the Bunch-Davies vacuum in a unitary QFT in dS$_2$ has a K\"allén-Lehmann decomposition into UIRs of $SO(1,2)$ as follows\footnote{In our conventions $G_\Delta$ is dimensionless, so the mass dimensions of the trace of the stress tensor are captured by the spectral densities, which thus have mass dimension 4.}\cite{Loparco:2023rug,DiPietro:2021sjt,Hogervorst:2021uvp,Bros:1990cu,Bros:1995js,Bros:1998ik,Bros:2009bz,Bros:2010rku}
\begin{equation}
    G_\Theta(\sigma)=2\pi\int_{\frac{1}{2}+i\mathbb{R}}\frac{d\Delta}{2\pi i}\ \varrho_\Theta^{\mathcal{P}}(\Delta)G_\Delta(\sigma)+\int_0 ^1d\Delta\ \varrho_\Theta^{\mathcal{C}}(\Delta)G_\Delta(\sigma)\,,
    \label{eq:KLtheta}
\end{equation}
with
\begin{equation}
    G_\Delta(\sigma)=\frac{1}{4}\csc(\pi\Delta)\ _2F_1\left(\Delta,\bar\Delta,1,\frac{1+\sigma}{2}\right)\,, \qquad \bar\Delta\equiv1-\Delta\,,
    \label{eq:freeprop}
\end{equation}
where $\Delta$ parametrizes the eigenvalue of the quadratic Casimir of $SO(1,2)$ as follows
\begin{equation}
    \mathcal{C}_2^{SO(1,2)}=\Delta(1-\Delta)\equiv m^2R^2\,.
\end{equation}
The first term in (\ref{eq:KLtheta}) stands for contributions associated to principal series UIRs, while the second one stands for complementary series contributions. In a unitary theory, the spectral densities $\varrho^{\mathcal{P}}_\Theta(\Delta)$ and $\varrho^{\mathcal{C}}_\Theta(\Delta)$ are positive on their domains of integration. For early references on the full classification of UIRs of $SO(1,2)$ see \cite{Dirac:1945cm,Bargmann:1946me,9fea073f-7ff0-3fb5-984e-cb6ac72836f7,MR0024440}. For recent reviews, see \cite{Sun:2021thf,Kitaev:2017hnr}.

Notice that we are excluding the possibility that, in two dimensions, the discrete series of UIRs ($\Delta=p\in\mathbb{N}\setminus\{0\}$) could contribute to the K\"allén-Lehmann decomposition of the trace of the stress tensor, since it is a scalar operator. In \cite{Loparco:2023akg,Loparco:2023rug} we phrased more precisely some arguments in favor of the fact that only operators with spin $J\geq p$ can interpolate between the vacuum and states in the discrete series. 
The rigorous statement is the following: the two-point function of an operator with $J$ indices which does not grow at infinite geodesic distance and does not have branch cuts at spacelike separation, cannot have discrete series irreps with $p>J$ in its K\"allén-Lehmann decomposition. In the $J=0$ case, the reason is that the Casimir equation $\left(\Box-p(1-p)\right)G_p=0$ has two families of solutions, one which grows at infinity and one which has a branch point at $\sigma=-1$. Concretely, the solutions with the unphysical cuts are Legendre Q functions $Q_p(\sigma)$. Taking a discontinuity across the branch cuts returns Legendre P functions $P_p(\sigma)$, which are orthogonal for different $p$. Thus, there is no way that a finite sum of discrete series irreps leads to a two-point function which is free of unphysical branch cuts. Similar statements can be adapted to operators with $J$ indices, see \cite{Loparco:2023akg,Loparco:2023rug} for more details. Then, we conjecture that it is impossible that an infinite sum of discrete series contributions leads to a two-point function which is free of unphysical branch cuts. This is not an obvious statement because the infinite sum and the discontinuity do not commute.

Some references which further speculate on possible loopholes to these arguments are \cite{Anninos:2023lin,Bros:2010wa,Epstein:2014jaa}. At the moment, no explicit counterexample to this conjecture is present in the literature.

Given these assumptions, the derivation of the spectral sum rule is straightforward: we plug (\ref{eq:KLtheta}) into (\ref{eq:sumrule2}) and carry out the integral over $\sigma$. We use the following identities
\begin{equation}
\begin{aligned}
    \int_{-1}^1d\sigma\ G_\Delta(\sigma)&=\frac{1}{2\pi}\frac{1}{\Delta\bar\Delta}\,,\\
    \int_{-1}^1d\sigma\ \sigma\ G_\Delta(\sigma)&=\frac{1}{2\pi}\frac{1}{(\Delta+1)(\bar\Delta+1)}\,,\\
    \int_{-1}^1d\sigma\ \sigma\log\left(\frac{1+\sigma}{2}\right) G_\Delta(\sigma)&=\frac{1}{2\pi}\frac{\Delta\bar\Delta-4}{\Delta\bar\Delta(\Delta+1)^2(\bar\Delta+1)^2}-\frac{\text{csc}(\pi\Delta)}{2(\Delta+1)(\bar\Delta+1)}\,.
\end{aligned}
\end{equation}
We obtain
\begin{shaded}
\begin{equation}
\begin{aligned}
	c^{\text{UV}}-c_1(R)=&\int_{\frac{1}{2}+i\mathbb{R}}\frac{d\Delta}{2\pi i}\left(\frac{24\pi^2}{(\Delta+1)^2(\bar\Delta+1)^2}+\frac{8\pi^3\csc(\pi\Delta)}{(\Delta+1)(\bar\Delta+1)}\right)R^4\varrho^{\mathcal{P}}_\Theta(\Delta)\\
 &+\int_0^1 \frac{d\Delta}{2\pi}\left(\frac{24\pi^2}{(\Delta+1)^2(\bar\Delta+1)^2}+\frac{8\pi^3\csc(\pi\Delta)}{(\Delta+1)(\bar\Delta+1)}\right)R^4\varrho ^{\mathcal{C}}_\Theta(\Delta)\,.
\label{eq:sumrule3}
\end{aligned}
\end{equation}
\end{shaded}
\noindent Notice that, now, the integrands on the right hand side are manifestly positive on the principal and complementary series domains $\Delta=\frac{1}{2}+i\mathbb{R}$ and $\Delta\in(0,1)$, implying that 
\begin{shaded}
\begin{equation}
    c^{\text{UV}}\geq c_1(R)\,.
\end{equation}
\end{shaded}
\paragraph{Flat space limit} Let us show that (\ref{eq:sumrule3}) reduces to (\ref{eq:flatdeltac}) in the flat space limit. We start from the fact that, reinstating factors of the radius, \cite{Loparco:2023rug} 
\begin{equation}
    \lim_{R\to\infty}\frac{R}{\sqrt{s}}\varrho_\mathcal{O}^{\mathcal{P}}(\Delta=iR\sqrt{s})=\varrho_{\mathcal{O}}^{\text{flat}}(s)\,,
\end{equation}
where $s\equiv m^2$ is the flat space mass that is integrated over in the K\"allén-Lehmann representation. Taking the flat space limit of (\ref{eq:sumrule3}) gives
\begin{equation}
\begin{aligned}
    c^{\text{UV}}-\lim_{R\to\infty}c_1(R)&=12\pi\lim_{R\to\infty}\left[\int_{0 }^{\infty}\frac{Rds}{\sqrt{s}}\frac{1}{s^2R^4}\frac{\sqrt{s}}{R}R^4\varrho_\Theta^{\text{flat}}(s)+\text{complementary}\right]\\
&=12\pi\int_{0}^\infty \frac{ds}{s^2}\varrho^{\text{flat}}_\Theta(s)+\lim_{R\to\infty}\text{complementary}\\
&=c^{\text{UV}}-c^{\text{IR}}+\lim_{R\to\infty}\text{complementary}\,,
\end{aligned}
\end{equation}
where we used the fact that the csc$(\pi\Delta)$ factor exponentially suppresses the spectral density in this limit, which cannot compete due to Tauberian theorems in flat space, and that the first term in the penultimate line was exactly the spectral sum rule (\ref{eq:flatdeltac}). Since we showed that $\lim_{R\to\infty}c_1(R)=c^{\text{IR}}$ in the previous section, we just proved that
\begin{equation}
    \lim_{R\to\infty}\int_0^1 \frac{d\Delta}{2\pi}\left(\frac{24\pi^2}{(\Delta+1)^2(\bar\Delta+1)^2}+\frac{8\pi^3\csc(\pi\Delta)}{(\Delta+1)(\bar\Delta+1)}\right)R^4\varrho ^{\mathcal{C}}_\Theta(\Delta)=0\,,
    \label{eq:vanishingcompl}
\end{equation}
meaning that the complementary series contribution has to vanish when taking the $R\to\infty$ limit.
\subsection{Behavior of the first c-function at vanishing radius}
\label{subsec:r0vanish}
Let us study the behavior of $c_1(R)$ as we take $R\to0$. We start from (\ref{eq:sumrule2}) and study the limit 
\begin{equation}
    \lim_{R\to0}\int_{-1}^1 d\sigma\left[1-\sigma\left(\log\left(\frac{1+\sigma}{2}\right)+1\right)\right]\tilde G_{\Theta}(\sigma,\{m_iR\})\,,
\end{equation}
where $\tilde G_\Theta\equiv R^4G_\Theta$ is the dimensionless two-point function of the trace of the stress tensor, and we made explicit the fact that it can in general depend on all the dimensionless combinations of the mass scales of the theory and of the radius. Taking $R\to0$ in this formula while keeping $m_k$ fixed is equivalent to taking all $m_k\to0$ and keeping $R$ fixed, probing the UV fixed point of the theory, which is a CFT on a two-sphere of radius $R$. It is a general fact that in a CFT in curved space, the two-point function of the trace of the stress tensor vanishes up to contact terms\footnote{The precise expression is $\langle\Theta(x_1)\Theta(x_2)\rangle=-\frac{c}{12\pi}\nabla^2\delta^{(2)}(x_1-x_2)$ \cite{Hartman:2023ccw}}. Then, we notice two more facts: the divergence of the kernel at $\sigma=-1$ is logarithmic and thus integrable, and the divergence of $\tilde G_\Theta$ at $\sigma=1$ is logarithmic and cured by a simple zero in the kernel. We can thus safely state that
\begin{equation}
    \lim_{R\to0}c_1(R)=c^{\text{UV}}\,, \qquad \text{if}\quad \lim_{R\to0}\tilde G_\Theta(\sigma)=0\,.
    \label{cuv10}
\end{equation}
This has implications regarding the integrals appearing in the spectral sum rule (\ref{eq:sumrule3}). Specifically, we can say that necessarily
\begin{equation}
\begin{aligned}
    \lim_{R\to0}\Big[&\int_{\frac{1}{2}+i\mathbb{R}}\frac{d\Delta}{2\pi i}\left(\frac{24\pi^2}{(\Delta+1)^2(\bar\Delta+1)^2}+\frac{8\pi^3\csc(\pi\Delta)}{(\Delta+1)(\bar\Delta+1)}\right)R^4\varrho^{\mathcal{P}}_\Theta(\Delta)\\
 &+\int_0^1 \frac{d\Delta}{2\pi}\left(\frac{24\pi^2}{(\Delta+1)^2(\bar\Delta+1)^2}+\frac{8\pi^3\csc(\pi\Delta)}{(\Delta+1)(\bar\Delta+1)}\right)R^4\varrho ^{\mathcal{C}}_\Theta(\Delta)\Big]=0\,.
 \label{eq:vanish2r0}
 \end{aligned}
\end{equation}
under the same assumption as in (\ref{cuv10}). This will be important when studying the second c-function in the next section. 
\paragraph{An important caveat}
Let us discuss the generality of the hypothesis in equation (\ref{cuv10}). While it is certainly true that $\tilde G_\Theta(\sigma)=0$ in a CFT on $S^2$, at least one case is known of an RG flow in which $\lim_{R\to0}\tilde G_\Theta(\sigma)\neq0$: the free massive boson flow. The cause of this can be traced back to the fact that the massless limit of the two-point function of a free field is divergent, a fact which itself originates from the IR divergent zero mode of the free massless scalar \cite{LopezNacir:2016gfj,Rajaraman:2010xd}. More details on this will be explained in Section \ref{subsec:freescalarmain}. The main statement is that there exist QFTs which violate the hypothesis $\lim_{R\to0}\tilde G_\Theta(\sigma)=0$, in which case $c_1(R)$ does not reach $c^{\text{UV}}$ as $R\to0$. Intuition suggests that, more in general, the UV of the theory must be ill defined in some sense in order to have $\lim_{R\to0}\tilde G_\Theta(\sigma)\neq0$. We will outline a technical hypothesis under which $c_2(R)$ interpolates to $c^{\text{UV}}$ in Section \ref{subsec:secondcfunc}, and see that a broader class of theories satisfies it, including the free massive boson.
\section{Spectral densities of the stress tensor and the second c-function}
\label{sec:stresstensor}
In this section, we derive a series of properties regarding the spectral representation of the stress tensor in unitary QFTs in de Sitter and on the sphere. In subsection \ref{subsec:spectral1}, we show the most general form of the spectral decomposition of the stress tensor in $d+1$ dimensions, taking into consideration only the contributions from the principal and complementary series. The conservation of the stress tensor implies relations between the spectral densities associated to different $SO(d)$ spin, making the expressions much simpler than initially expected. In subsection \ref{subsec:spectral2}, we specify to the case of two dimensions and we take into account all the UIRs that the stress tensor can in principle couple to. After imposing conservation, we find once again a much more compact expression than one would expect, and we show that from the discrete series only the $\Delta=2$ UIR can appear. We leave many details of this section to the Appendix \ref{subsec:spectralrelations}.

\subsection{Spectral decomposition of the stress tensor in higher dimensions}
\label{subsec:spectral1}
The stress tensor is a symmetric spin 2 operator. As such, \emph{naively}, one would expect it to have a total of five independent spectral densities: one associated to its trace, three associated to the $SO(d)$ decomposition of its traceless part, and one associated to the mixed two-point function of its trace and its traceless parts. In equations,
    \begin{align}
         \langle T^{AB}(Y_1)T^{CD}(Y_2)\rangle=2\pi\int_{\frac{d}{2}+i\mathbb{R}}\frac{d\Delta}{2\pi i}\Bigg[&\sum_{\ell=0}^2\varrho_{\hat T,\ell}^{\mathcal{P}}(\Delta)G^{AB,CD}_{\Delta,\ell}(Y_1,Y_2)\nonumber\\
         &+\varrho^{\mathcal{P}}_{\hat T\Theta}(\Delta)\left(\frac{G_2^{CD}}{d+1}\hat\Pi_1^{AB}G_{\Delta}(\sigma)+\frac{G_1^{AB}}{d+1}\hat\Pi_2^{CD}G_{\Delta}(\sigma)\right)\nonumber\\
    &+\varrho^{\mathcal{P}}_\Theta(\Delta)\frac{G_1^{AB}G_2^{CD}}{(d+1)^2}G_\Delta(\sigma)\Bigg]+\text{other UIRs}\,.
    \label{eq:KLstress}
    \end{align}
where $G^{AB,CD}_{\Delta,\ell}(Y_1,Y_2)$ are the blocks that appear in the K\"allén-Lehmann representation of traceless symmetric spin 2 operators on the Bunch-Davies vacuum, $\hat\Pi_i^{AB}$ is a traceless symmetric differential operator
\begin{equation}
    \hat\Pi_i^{AB}=\frac{1}{d+1}G_i^{AB}\nabla_i ^2-\nabla_i^{(A}\nabla_i^{B)}\,,
    \label{eq:pihatdef}
\end{equation}
and $G_\Delta(\sigma)$ is the canonically normalized free scalar propagator in $d+1$ dimensions in the Bunch-Davies vacuum \cite{Spradlin:2001pw}\footnote{We use the notation for the regularized hypergeometric function $\mathbf{F}(a,b,c,z)\equiv\frac{1}{\Gamma(c)}\ _2F_1(a,b,c,z)\,,$ and in this section $\bar\Delta\equiv d-\Delta$, while in the rest of the paper $\bar\Delta\equiv 1-\Delta$.}
\begin{equation}
	G_\Delta(\sigma)=\frac{\Gamma(\Delta)\Gamma(\bar\Delta)}{(4\pi)^{\frac{d+1}{2}}}\mathbf{F}\left(\Delta,\bar\Delta,\frac{d+1}{2},\frac{1+\sigma}{2}\right)\,, \qquad \sigma=\frac{1}{R^2}Y_1\cdot Y_2\,.
\label{eq:freepropd1}
\end{equation}
In particular, the $\ell=2$ block $G^{AB,CD}_{\Delta,2}(Y_1,Y_2)$ is explicitly reported in (\ref{eq:Gdelta2exp}) and the ones for $\ell=0$ and $\ell=1$ can be found in index-free form in appendix F.3 of \cite{Loparco:2023rug}. Embedding space covariant derivatives $\nabla^A$ and the induced metric $G^{AB}$ are defined in Section \ref{subsec:preliminaries}. In this section, we are again setting $R=1$.

Taking or removing traces from (\ref{eq:KLstress}) reduces it to the following, naively independent, decompositions
\begin{equation}
\begin{aligned}
    \langle\hat T^{AB}(Y_1)\hat T^{CD}(Y_2)\rangle&=2\pi\int_{\frac{d}{2}+i\mathbb{R}}\frac{d\Delta}{2\pi i}\sum_{\ell=0}^2\varrho_{\hat T,\ell}^{\mathcal{P}}(\Delta)G^{AB,CD}_{\Delta,\ell}(Y_1,Y_2)+\text{other UIRs}\,,\\
    \langle\Theta(Y_1)\hat T^{CD}(Y_2)\rangle&=2\pi\int_{\frac{d}{2}+i\mathbb{R}}\frac{d\Delta}{2\pi i}\ \varrho_{\hat T\Theta}^{\mathcal{P}}(\Delta)\hat \Pi^{CD}_2G_\Delta(\sigma)+\text{other UIRs}\,,\\
    \langle\Theta(Y_1)\Theta(Y_2)\rangle&=2\pi\int_{\frac{d}{2}+i\mathbb{R}}\frac{d\Delta}{2\pi i}\ \varrho^{\mathcal{P}}_\Theta(\Delta)G_\Delta(\sigma)+\text{other UIRs}\,.
\end{aligned}
\end{equation}
The conservation of the stress tensor induces relations among these spectral densities, totally analogous to those in flat space \cite{Karateev:2020axc}. We relegate the proof of these relations to Appendix \ref{subsec:spectralrelations}. Here, we report the results\footnote{Here we omit the superscripts on the spectral densities specifying the series of UIRs because these identities apply also to the complementary series, given that the functional form of its contribution is just the analytic continuation of the principal series ones.}
\begin{equation}
    \varrho_{\hat T\Theta}(\Delta)=\frac{\varrho_\Theta(\Delta)}{d(\Delta+1)(\bar\Delta+1)}\,, \quad \varrho_{\hat T,0}(\Delta)=\frac{\varrho_\Theta(\Delta)}{d^2(\Delta+1)^2(\bar\Delta+1)^2}\,,\quad \varrho_{\hat T,1}(\Delta)=0\,.
    \label{eq:spectralrelations}
\end{equation}
The K\"allén-Lehmann decomposition of the stress tensor thus reduces to
\begin{shaded}
\begin{equation}
\begin{aligned}
    \langle T^{AB}(Y_1)T^{CD}(Y_2)\rangle=2\pi\int_{\frac{d}{2}+i\mathbb{R}}\frac{d\Delta}{2\pi i}\Big[&\varrho^{\mathcal{P}}_{\hat T,2}(\Delta)G^{AB,CD}_{\Delta,2}(Y_1,Y_2)\\
    &+\frac{\varrho^{\mathcal{P}}_\Theta(\Delta)}{d^2(\Delta+1)^2(\bar\Delta+1)^2}\Pi_1^{AB}\Pi_2^{CD}G_\Delta(\sigma)\Big]\\
    &+\text{other UIRs}
    \label{eq:stresstensorspectral}
\end{aligned}
\end{equation}
\end{shaded}
\noindent where $\Pi_i^{AB}$ comes from the combination of the various propagators proportional to $\varrho_\Theta$ after applying (\ref{eq:spectralrelations})
\begin{equation}
\Pi_i^{AB}\equiv G_i^{AB}\left(d+\nabla_i^2\right)-\nabla_i^{(A}\nabla_i^{B)}\,.
\label{eq:definepi}
\end{equation}
In this representation, both lines in (\ref{eq:stresstensorspectral}) are independently conserved: $\nabla_AG_{\Delta,2}^{AB,CD}=0$ by definition and it can be checked that $\nabla_A\Pi^{AB}G_\Delta=0$. Group theoretically, the first line corresponds to states which carry $SO(d)$ spin 2, while the second line corresponds to all other scalar states. 

\subsection{Spectral decomposition of the stress tensor in two dimensions}
\label{subsec:spectral2}
In two dimensions, the picture simplifies even further: there is no dynamical propagating massive traceless symmetric spin 2 field, so $G^{AB,CD}_\Delta(Y_1,Y_2)=0$. Moreover, the only UIRs that can contribute, other than the principal series, are the complementary series and the irrep with $\Delta=2$ in the discrete series. We prove this fact in Appendix \ref{subsec:spectralrelations}. We are left with\footnote{We use the same notation for projectors and propagators that we used in the higher dimensional case, but we are implicitly setting $d=1$.}
\begin{shaded}
\begin{equation}
\begin{aligned}
    \langle T^{AB}(Y_1)T^{CD}(Y_2)\rangle=&2\pi\int_{\frac{1}{2}+i\mathbb{R}}\frac{d\Delta}{2\pi i}\frac{\varrho^{\mathcal{P}}_\Theta(\Delta)}{(\Delta+1)^2(\bar\Delta+1)^2}\Pi_1^{AB}\Pi_2^{CD}G_\Delta(\sigma)\\
    &+\int_0^1 d\Delta\frac{\varrho^{\mathcal{C}}_\Theta(\Delta)}{(\Delta+1)^2(\bar\Delta+1)^2}\Pi_1^{AB}\Pi_2^{CD}G_\Delta(\sigma)\\
    &+\varrho^{\mathcal{D}_2}_{\hat T}\Pi_1^{AB}\Pi_2^{CD}G_{\Delta=2}(\sigma)\,.
    \label{eq:stresstensorspectral2d}
\end{aligned}
\end{equation}
\end{shaded}
\noindent Since for the discrete series there is no integral over $\Delta$, we call $\varrho^{\mathcal{D}_2}_{\hat T}$ the spectral weight of the stress tensor in the $\Delta=2$ UIR. 

When the theory has a good continuation in the number of spacetime dimensions, there is a final simplification. If in higher dimensions the stress tensor only decomposes in principal and complementary series, then when continuing to $d=1$ the only contribution to the $\Delta=2$ discrete series comes from spurious poles at $\Delta=2$ and $\bar\Delta=2$ in $G_{\Delta,2}^{AB,CD}$ which will cross the contour of integration over the principal series and lead to the discrete series $\Delta=2$ contribution. This is in fact what happens in the free massive boson case, as we will discuss in further detail in Section \ref{subsec:freescalarmain} and Appendix \ref{subsec:detailsboson}. We can thus state that if the theory has a good analytic continuation in $d$, with only principal and complementary series contributions to the stress tensor in higher dimensions, we have
\begin{equation}
    \varrho^{\mathcal{D}_2}_{\hat T}=4\pi\ \underset{\Delta=2}{\text{Res}}\left(\frac{\varrho^{\mathcal{P}}_\Theta(\Delta)}{(\Delta+1)^2(\bar\Delta+1)^2}\right)\,.
\end{equation}
\subsection{Finding the second c-function}
\label{subsec:secondcfunc}
By comparing the definitions (\ref{eq:twopointstructs}), (\ref{eq:TtotildeT}) and (\ref{eq:defcR}) with the spectral decomposition (\ref{eq:stresstensorspectral2d}), it is possible to derive formulas which extract $c^{\text{UV}}$ and $c_1(R)$ individually as integrals over the spectral densities of the stress tensor. To start, in Appendix \ref{subsec:Tifuncs} we show how to relate the $\mathcal{T}_i(\sigma)$ functions to integrals over the spectral densities of the stress tensor, obtaining equations (\ref{eq:TTTTspectrals}). Then, evaluating them at $\sigma=-1$, we obtain
\begin{equation}
\begin{aligned}
    &\mathcal{T}_1(-1)=0\,, \qquad \qquad\qquad\qquad\qquad\qquad\qquad\qquad\qquad\qquad\mathcal{T}_2(-1)=0\,, \\
    &\mathcal{T}_3(-1)=-\frac{3}{32\pi}\varrho_{\hat T}^{\mathcal{D}_2}+\frac{\pi}{32}\int_{\frac{1}{2}+i\mathbb{R}}\frac{d\Delta}{2\pi i}\frac{(4+\Delta\bar\Delta)\text{csc}(\pi\Delta)}{(\Delta+1)(\bar\Delta+1)}\varrho_\Theta^{\mathcal{P}}(\Delta)+\text{complementary}\,,\\
    &\mathcal{T}_4(-1)=\frac{3}{32\pi}\varrho^{\mathcal{D}_2}_{\hat T}+\frac{\pi}{32}\int_{\frac{1}{2}+i\mathbb{R}}\frac{d\Delta}{2\pi i}\frac{(4+3\Delta\bar\Delta)\text{csc}(\pi\Delta)}{(\Delta+1)(\bar\Delta+1)}\varrho_\Theta^{\mathcal{P}}(\Delta)+\text{complementary}\,,
\end{aligned}
\end{equation}
where ``complementary" stands for the same exact expression as the principal series case but with an integral over the $\Delta\in(0,1)$ contour.
Now, using the definition of $c_1(R)$ (\ref{eq:defcR}), we get
\begin{equation}
    c_1(R)=12\pi R^4\left(\varrho^{\mathcal{D}_2}_{\hat T}-\frac{2\pi^2}{3}\int_{\frac{1}{2}+i\mathbb{R}}\frac{d\Delta}{2\pi i}\frac{\text{csc}(\pi\Delta)\varrho_\Theta^{\mathcal{P}}(\Delta)}{(\Delta+1)(\bar\Delta+1)}-\frac{\pi}{3}\int_0^1 d\Delta\frac{\text{csc}(\pi\Delta)\varrho_\Theta^{\mathcal{C}}(\Delta)}{(\Delta+1)(\bar\Delta+1)}\right)\,.
    \label{eq:crsum}
\end{equation}
Using our sum rule (\ref{eq:sumrule3}), we can thus derive a formula for $c^{\text{UV}}$ which is valid for any $R$:
\begin{equation}
       c^{\text{UV}}=12\pi R^4\left(\varrho^{\mathcal{D}_2}_{\hat T}+2\pi\int_{\frac{1}{2}+i\mathbb{R}}\frac{d\Delta}{2\pi i}\frac{\varrho_\Theta^{\mathcal{P}}(\Delta)}{(\Delta+1)^2(\bar\Delta+1)^2}+\int_0^1 d\Delta\frac{\varrho_\Theta^{\mathcal{C}}(\Delta)}{(\Delta+1)^2(\bar\Delta+1)^2}\right)\,.
    \label{eq:cuvsum}
\end{equation}
Interestingly, in the flat space limit the second term in (\ref{eq:cuvsum}) independently reduces to the sum rule for $c^{\text{UV}}-c^{\text{IR}}$, see the previous paragraph. At the same time, the principal series integral in (\ref{eq:crsum}) vanishes in this limit. Moreover, (\ref{eq:vanishingcompl}) implies that both complementary series integrals in (\ref{eq:cuvsum}) and (\ref{eq:crsum}) vanish in this limit. Finally, in Section \ref{subsec:r0vanish} we showed that all of these integrals of the spectral densities of the trace of the stress tensor vanish as $R\to0$ if $\lim_{R\to0}G_\Theta(\sigma)=0$. We can thus define 
\begin{shaded}
\begin{equation}
    c_2(R)\equiv 12\pi R^4\varrho^{\mathcal{D}_2}_{\hat T}\,,
    \label{eq:defc2r}
\end{equation}
\end{shaded}
\noindent and state that
\begin{equation}
\begin{aligned}
    \lim_{R\to\infty}c_2(R)&=c^{\text{IR}}\,,\\
    \lim_{R\to0}c_2(R)&=c^{\text{UV}}\,, \qquad \text{if}\quad \lim_{R\to0}G_\Theta(\sigma)=0\,.
    \label{eq:cirdelta2}
\end{aligned}
\end{equation}
\noindent In other words, the spectral weight of the stress tensor in the discrete series $\Delta=2$ irrep is another candidate $c$-function which interpolates between $c^{\text{IR}}$ and $c^{\text{UV}}$ as we vary the radius. In the last part of this section we will weaken the assumption in (\ref{eq:cirdelta2}).

\begin{figure}
\centering
\includegraphics[scale=1.3]{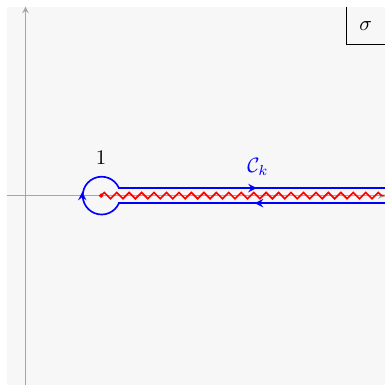}
\caption{In blue, the contour of integration $\mathcal{C}_k$ in (\ref{eq:sphereinverse}). It wraps the branch cut of the two-point function for time-like separation. In practice, it is equivalent to summing the residue at $\sigma=1$ and the discontinuity at $\sigma\in(1,\infty)$.}
\label{fig:keyhole}
\end{figure}

Let us write down two sum rules for $c_2(R)$. The one in terms of spectral densities is obtained by combining (\ref{eq:crsum}), (\ref{eq:cuvsum}) and (\ref{eq:defc2r}):
\begin{shaded}
\begin{equation}
    c^{\text{UV}}-c_2(R)=24\pi^2 R^4\left(\int_{\frac{1}{2}+i\mathbb{R}}\frac{d\Delta}{2\pi i}\frac{\varrho_\Theta^{\mathcal{P}}(\Delta)}{(\Delta+1)^2(\bar\Delta+1)^2}+\int_0^1 \frac{d\Delta}{2\pi}\frac{\varrho_\Theta^{\mathcal{C}}(\Delta)}{(\Delta+1)^2(\bar\Delta+1)^2}\right)
    \label{eq:newrule}
\end{equation}
\end{shaded}
\noindent
once again from this we can deduce $c^{\text{UV}}\geq c^{\text{IR}}$. Comparing with (\ref{eq:sumrule3}) we can  further state
\begin{equation}
    c_2(R)\geq c_1(R)\,. \label{eq:c2gc1}
\end{equation}
Deriving the position space sum rule for $c_2(R)$ is slightly more involved. We make use of the inversion formula from \cite{Hogervorst:2021uvp,Chakraborty:2023qbp,Loparco:2023rug}, which in two dimensions states that the principal series spectral density associated to a two-point function $G(\sigma)$ is given by
\begin{equation}
    \rho^{\mathcal{P}}(\Delta)=\left(\frac{1}{2}-\Delta\right)i\cot(\pi\Delta)\int_{\mathcal{C}_k}d\sigma\ _2F_1\left(\Delta,\bar\Delta,1,\frac{1-\sigma}{2}\right)G(\sigma)\,,
    \label{eq:sphereinverse}
\end{equation}
with the contour $\mathcal{C}_k$ being a ``keyhole" contour wrapping the branch cut of $G(\sigma)$, which for a physical two-point function is at $\sigma\in[1,\infty)$, see figure \ref{fig:keyhole}. In practice, evaluating this integral corresponds to computing the residue of the integrand at $\sigma=1$ and the discontinuity of $G(\sigma)$ around the cut.

For now, we will assume that there are no further contributions to the spectral decomposition of our two-point function. We will see that the sum rule we obtain in this way works even if there are complementary series contributions.

To proceed, we plug (\ref{eq:sphereinverse}) inside (\ref{eq:newrule}), and we use the following identity 
\begin{equation}
    _2F_1\left(\Delta,\bar\Delta,1,\frac{1-\sigma}{2}\right)=\frac{\Gamma(\frac{1}{2}-\Delta)}{\sqrt{\pi}\Gamma(\bar\Delta)}\frac{1}{(2(1+\sigma))^{\Delta}}\ _2F_1\left(\Delta,\Delta,2\Delta,\frac{2}{1+\sigma}\right)+(\Delta\leftrightarrow\bar\Delta)\,.
    \label{eq:2F1ident}
\end{equation}
Exploiting the symmetry of the integral, we can drop the second term in (\ref{eq:2F1ident}). Now the integrand decays with large $\text{Re}(\Delta)$. Assuming the convergence of the sum rule (\ref{eq:newrule}) and of the inversion formula (\ref{eq:sphereinverse}), we can swap the integrals and close the contour of integration over $\Delta$ on the right half of the complex plane, picking up the residue on the only pole of the integrand, which is at $\Delta=2$. We obtain the following position space sum rule,
\begin{shaded}
\begin{equation}
    c^{\text{UV}}-c_2(R)=\int_{\mathcal{C}_k} d\sigma\ r_2(\sigma)R^4G_{\Theta}(\sigma)\,.
    \label{eq:newsumrule}
\end{equation}
\end{shaded}
\noindent
with the explicit form of the kernel being
\begin{align}
  r_2(\sigma)=&\frac{4\pi i}{(1+\sigma)^2}\Bigg[
    2\sigma ^3\text{coth}^{-1}(\sigma)+\sigma ^2 \log \left(\frac{(\sigma +1)^3}{4(\sigma-1)}\right)+\sigma \log \left(\frac{(\sigma -1)(\sigma+1)^3}{16}\right)\\
    &+\log \left(\frac{\sigma ^2-1}{4}\right)+2(1+\sigma)^2\left(1-\sigma\left(\text{coth}^{-1}(\sigma)\log\left(\frac{\sigma-1}{2}\right)-\text{Li}_2\left(\frac{2}{1-\sigma}\right)\right)\right)\Bigg]\nonumber\label{eq:r2explicit}\,,
\end{align}
where $\text{Li}_2(x)$ is a dilogarithm. Notice that $r_2(\sigma)$ is purely imaginary in $\sigma\in(1,\infty)$, as is the discontinuity of $G_{\Theta}(\sigma)$, so that the integrand in (\ref{eq:newsumrule}) is real. 

Given this new sum rule (\ref{eq:newsumrule}) we can now weaken the assumption in (\ref{eq:cirdelta2}). Let us write explicitly
\begin{equation}
    \int_{\mathcal{C}_k}d\sigma\  r_2(\sigma)\tilde G_\Theta(\sigma)=\int_1^\infty d\sigma\ r_2(\sigma)\text{Disc}[\tilde G_\Theta(\sigma)]-2\pi ir_2(1)\underset{\sigma=1}{\text{Res}}[\tilde G_\Theta(\sigma)]\,,
\end{equation}
where we are using the notation of Section \ref{subsec:r0vanish} where $\tilde G_\Theta\equiv R^4G_\Theta$. From this, we can state the following sufficient condition
\begin{equation}
    \lim_{R\to0}c_2(R)=c^{\text{UV}}\qquad \text{if}\quad \lim_{R\to0}\text{Disc}[\tilde G_\Theta(\sigma)]\quad \text{and}\quad \lim_{R\to0}\underset{\sigma=1}{\text{Res}}[\tilde G_\Theta(\sigma)]=0\,.
    \label{eq:criterc2}
\end{equation}
This is a more general condition than the one for $c_1(R)$ (\ref{cuv10}). Infact, in the free boson case, where $\lim_{R\to0}c_1(R)\neq c^{\text{UV}}$, we instead have $\lim_{R\to0}c_2(R)=c^{\text{UV}}$. The fact that this condition is weaker puts $c_2(R)$ on a preferred standing with respect to $c_1(R)$ as a candidate c-function.

For completeness, let us state that as long as the two-point function $G_\Theta(\sigma)$ can be analytically continued to some regime where only the principal series contributes to its spectral decomposition, then (\ref{eq:newsumrule}) works in every other regime. We checked that this works in the free massive boson case, even when complementary series contributions appear. 

As a final note, let us emphasize that what we showed in this section implies that in any QFT\footnote{The only exception is, of course, the empty theory.} the spectral decomposition of the stress tensor \emph{must} contain a contribution from the $\Delta=2$ discrete series irrep, since its spectral weight has to interpolate between $c^{\text{IR}}$ and $c^{\text{UV}}$.
\subsection{Independent argument for the second c-function}
Here we will give an independent argument for why $\varrho^{\mathcal{D}_2}_{\hat T}$ interpolates between $c^{\text{UV}}$ and $c^{\text{IR}}$ as we tune the radius of the sphere $R$. Let us start by writing down the K\"allén-Lehmann decomposition of the stress tensor in two-dimensional flat space \cite{Cappelli:1990yc,Karateev:2019ymz,Karateev:2020axc}
\begin{equation}
    \langle T^{\mu\nu}(x_1)T^{\rho\sigma}(x_2)\rangle^{\text{flat}}=\frac{c^{\text{IR}}}{12\pi}\Pi^{\mu\nu}_1\Pi^{\rho\sigma}_2G_0(x_1,x_2)+\int_0^\infty \frac{ds}{s^2}\tilde\varrho_\Theta(s)\Pi^{\mu\nu}_1\Pi^{\rho\sigma}_2G_{s}(x_1,x_2)\,,
    \label{eq:flatTspectral1}
\end{equation}
where we separated the massless contributions from the massive ones, and 
\begin{equation}
    G_s(x_1,x_2)\equiv\frac{1}{2\pi}K_0(\sqrt{s}|x_1-x_2|)\,,
\end{equation}
is the canonically normalized propagator of a massive free scalar with $m^2=s$ in two dimensions, with $K_n(x)$ being the modified Bessel function of the second kind, and 
\begin{equation}
    \Pi^{\mu\nu}_i\equiv \eta^{\mu\nu}\partial^2_i-\partial^\mu_i\partial^\nu_i\,,
\end{equation}
are the divergence-less projectors which ensure conservation of the stress tensor. Notice that the massless contribution in (\ref{eq:flatTspectral1}) is also traceless. That is necessary, since it is what survives in the IR CFT. In fact, it can be checked that 
\begin{equation}
    \frac{c^{\text{IR}}}{12\pi}\Pi^{\mu\nu}_1\Pi^{\rho\sigma}_2G_0(x_1,x_2)=\langle T^{\mu\nu}(x_1)T^{\rho\sigma}(x_2)\rangle_{\text{CFT}}^{\text{flat}}\,.
\end{equation}
On the other hand, consider the K\"allén-Lehmann decomposition of the stress tensor in $S^2/$dS$_2$ which we derived in the previous section and which we report here for convenience
\begin{equation}
    \begin{aligned}
    \langle T^{AB}(Y_1)T^{CD}(Y_2)\rangle=&2\pi\int_{\frac{1}{2}+i\mathbb{R}}\frac{d\Delta}{2\pi i}\frac{\varrho^{\mathcal{P}}_\Theta(\Delta)}{(\Delta+1)^2(\bar\Delta+1)^2}\Pi_1^{AB}\Pi_2^{CD}G_\Delta(\sigma)\\
    &+\int_0^1 d\Delta\frac{\varrho^{\mathcal{C}}_\Theta(\Delta)}{(\Delta+1)^2(\bar\Delta+1)^2}\Pi_1^{AB}\Pi_2^{CD}G_\Delta(\sigma)\\
    &+\varrho^{\mathcal{D}_2}_{\hat T}\Pi_1^{AB}\Pi_2^{CD}G_{\Delta=2}(\sigma)\,.
    \label{eq:mmmm}
\end{aligned}
\end{equation}
In \cite{Loparco:2023rug} we studied the flat space limit of the principal series contributions and showed that they account for the continuum part in (\ref{eq:flatTspectral1}). Then, in (\ref{eq:vanishingcompl}) we argued that the complementary series part has to vanish in the flat space limit. What remains is only the last line in (\ref{eq:mmmm}). Now notice that the $\Delta=2$ contribution is precisely the two-point function of the stress tensor in a CFT on the two-sphere, up to a normalization factor
\begin{equation}
\begin{aligned}
    W^\pm_{1A}W^\pm_{1B}W^\pm_{2C}W^\pm_{2D}\Pi_1^{AB}\Pi_2^{CD}G_{\Delta=2}(\sigma)&=\frac{6}{\pi}\frac{(W_1^\pm\cdot W_2^\pm)^2}{(1-\sigma)^4}\\
    &\propto \langle T(Y_1,W_1^\pm)T(Y_2,W_2^\pm)\rangle_{\text{CFT}}^{\text{sphere}}\,,
    \label{eq:spherecft}
\end{aligned}
\end{equation}
where $W^\pm_{iA}$ are null vectors we are using to contract indices and give a compact form to the final expression, and the $\pm$ stands for their behavior under parity. We explain some more details on them in Appendix \ref{subsec:detailsboson} and in our previous work \cite{Loparco:2023rug}. Then, based on what we argued about the flat space limit, this is what matches the massless part in (\ref{eq:flatTspectral1}) when $R\to\infty$, so that 
\begin{equation}
    c^{\text{IR}}=12\pi \lim_{R\to\infty}R^4\varrho^{\mathcal{D}_2}_{\hat T}\,.
\end{equation}
On the other hand, as we discussed in \ref{subsec:r0vanish}, taking $R\to0$ is equivalent to probing the theory on the sphere at fixed radius but with all mass scales taken to zero, effectively flowing to the UV CFT on $S^2$/dS$_2$, where the spectral densities of the trace of the stress tensor vanish and the only term surviving in (\ref{eq:mmmm}) is the discrete series. Based on this physical assumption, let us assume that in this limit the integrals in (\ref{eq:mmmm}) vanish. Then,
\begin{equation}
    c^{\text{UV}}=12\pi \lim_{R\to0}R^4\varrho^{\mathcal{D}_2}_{\hat T}\,,
    \label{eq:cuvc2}
\end{equation}
giving an independent argument for why $c_2(R)$ defined in (\ref{eq:defc2r}) interpolates between $c^{\text{UV}}$ and $c^{\text{IR}}$. See the discussion around equation (\ref{eq:criterc2}) for a proof of a weaker assumption under which (\ref{eq:cuvc2}) is true.
\section{Examples}
\label{sec:examples}
In this section, we apply the sum rules (\ref{eq:sumrule2}), (\ref{eq:sumrule3}), (\ref{eq:newrule}) and (\ref{eq:newsumrule}) in the cases of a free massive scalar and a free massive fermion to compute the associated c-functions $c_1(R)$ and $c_2(R)$. In the free massive boson case we compute all the spectral densities of the stress tensor and show that the conservation relations (\ref{eq:spectralrelations}) are satisfied.
\subsection{Free massive scalar}
\label{subsec:freescalarmain}
Consider the theory of a free massive scalar with $m^2R^2=\Delta_\phi(1-\Delta_\phi)$. Without loss of generality, we will use the convention $\Delta_\phi=\frac{1}{2}\left(1+\sqrt{1-4(mR)^2}\right)$. The action is given by
\begin{equation}
    S=-\frac{1}{2}\int d^2x\sqrt{g}\left(g^{\mu\nu}\partial_\mu\phi\partial_\nu\phi+m^2\phi^2\right)\,,
    \label{eq:bosonaction}
\end{equation}
In the UV, this can be seen as the free theory of a massless scalar, for which we expect $c^{\text{UV}}=1$, perturbed by the relevant operator $m^2\phi^2$. Following the flow to the IR, we get to the trivial empty theory, $c^{\text{IR}}=0$. In flat space, this is one of the simplest examples of RG flows in QFT and the sum rules (\ref{eq:flatdeltac}) work perfectly fine. In de Sitter, the IR divergences associated to the zero mode of a massless scalar will instead slightly spoil this picture. 

As we take the radius to zero, in fact, we are going to find that the two-point function of the trace of the stress tensor becomes a non-zero constant, due to the divergence of $\langle\phi\phi\rangle$ in this limit, which is equivalent to the massless limit. This violates the condition stated in (\ref{cuv10}). Instead, the assumption in (\ref{eq:criterc2}) is not going to be violated, so $c_2(R)$ will successfully interpolate between $c^{\text{UV}}$ and $c^{\text{IR}}$. Let us discuss the details.

The stress tensor for this theory, computed from its definition (\ref{eq:defTmunu}), is
\begin{equation}
    T_{\mu\nu}=\partial_\mu\phi\partial_\nu\phi-\frac{1}{2}g_{\mu\nu}\left[\partial^\rho\phi\partial_\rho\phi+m^2\phi^2\right]\,.
    \label{eq:Tmunuboson}
\end{equation}
Its trace is $\Theta=-m^2\phi^2\,.$ The two-point function of the trace is thus 
\begin{equation}
G_\Theta(\sigma)=2m^4\left(G_{\Delta_\phi}(\sigma)\right)^2\,.
\label{eq:thetaboson}
\end{equation}
where $G_{\Delta}(\sigma)$ is in (\ref{eq:freeprop}).
As we take the radius to 0 fixing $m$, we get
\begin{equation}
    \lim_{R\to0}2m^4R^4\left(G_{\Delta_\phi}(\sigma)\right)^2=\frac{1}{8\pi^2}
\end{equation}
violating (\ref{cuv10}). Infact, if we just apply the sum rule (\ref{eq:defcR}) we get
\begin{equation}
    c_1(R)=0\,, \qquad \forall R
\end{equation}
implying that $c_1(R)=0$ for all $R$. In Appendix \ref{subsec:detailsboson}, we compute the full two-point function of $T^{\mu\nu}$ and independently verify that $c_1(R)=0$ using its definition (\ref{eq:defcR}). At the same time, we can check whether condition (\ref{eq:criterc2}) is satisfied. Using the following formulas for the discontinuity of hypergeometric functions
\begin{equation}
\begin{aligned}
    \text{Disc}\left[\mathbf{F}(a,b,c,z)\right]&=\frac{2\pi i}{\Gamma(a)\Gamma(b)}(z-1)^{c-a-b}\mathbf{F}\left(c-a,c-b,c-a-b+1,1-z\right)\,,\\
    \text{Disc}\left[\mathbf{F}^2(a,b,c,z)\right]&=\left(\text{Disc}[\mathbf{F}(a,b,c,z)]+2\mathbf{F}(a,b,c,z)\text{Disc}[\mathbf{F}(a,b,c,z)]\right)\,.
    \label{eq:discontinuities}
\end{aligned}
\end{equation}
one can easily check that criterion (\ref{eq:criterc2}) is verified. Then, we expect $c_2(R)$ to interpolate between $c^{\text{UV}}$ and $c^{\text{IR}}$ for this theory.

In order to derive an explicit expression for $c_2(R)$ we will use its definition in terms of the spectral density of the discrete series (\ref{eq:defc2r}). 
In Appendix \ref{subsec:detailsboson} we compute all the spectral densities of the stress tensor for this theory and we check formulas (\ref{eq:spectralrelations}) and (\ref{eq:sumrule3}). Let us report here the resulting decomposition. The general form of the spectral decomposition of the stress tensor was derived in Section \ref{sec:stresstensor}.
\begin{equation}
\begin{aligned}
    \langle T^{AB}(Y_1)T^{CD}(Y_2)\rangle=&2\pi\int_{\frac{1}{2}+i\mathbb{R}}\frac{d\Delta}{2\pi i}\frac{\varrho^{\mathcal{P}}_\Theta(\Delta)}{(\Delta+1)^2(\bar\Delta+1)^2}\Pi_1^{AB}\Pi_2^{CD}G_{\Delta}(\sigma)\\
    &+\int_0^1d\Delta\frac{\varrho^{\mathcal{C}}_\Theta(\Delta)}{(\Delta+1)^2(\bar\Delta+1)^2}\Pi_1^{AB}\Pi_2^{CD}G_{\Delta}(\sigma)\\
    &+\varrho^{\mathcal{D}_2}_{\hat T}\Pi_1^{AB}\Pi^{CD}_2G_{\Delta=2}(\sigma)\,,
\label{eq:spectralTT}
\end{aligned}
\end{equation}
where the differential operators $\Pi^{AB}_i$ where defined in (\ref{eq:definepi}). As expected from our arguments, we observe the presence of a discrete series irrep with $\Delta=2$. The explicit form of the spectral densities is, for the principal series
\begin{equation}
    \varrho^{\mathcal{P}}_\Theta\left(\frac{1}{2}+i\lambda\right)= \frac{m^4\lambda\sinh(\pi\lambda)}{16\pi^4\Gamma(\frac{1}{2}\pm i\lambda)}\Gamma\left(\frac{\frac{1}{2}\pm i\lambda}{2}\right)^2\prod_{\pm,\pm}\Gamma\left(\frac{\frac{1}{2}\pm i\lambda\pm 2i\lambda_\phi}{2}\right)\,,
\end{equation}
where we used $\Delta_\phi=\frac{1}{2}+i\lambda_\phi$ for convenience, so then $m^2 R^2=\frac{1}{4}+\lambda_\phi^2$. For the complementary and discrete series we find
\begin{align}
    \varrho^{\mathcal{C}}_\Theta(\Delta)&=-\delta(\Delta-2\Delta_\phi+1)\theta\left(\Delta_\phi-\frac{3}{4}\right)\frac{(\Delta+1)^2\bar\Delta\cos(\pi\Delta)\Gamma(\frac{3}{2}-\Delta)\Gamma(\frac{3-\Delta}{2})\Gamma(\frac{\Delta}{2})^2}{2^{4-\Delta}\pi^2R^4\Gamma(1-\frac{\Delta}{2})}\,,\nonumber\\
    \varrho^{\mathcal{D}_2}_{\hat T}&=\frac{\lambda_\phi m^2}{3R^2}\text{csch}(2\pi\lambda_\phi)\,,
    \label{eq:rhocrhodboson}
\end{align}
where we are using the convention $\Delta_\phi=\frac{1}{2}\left(1+\sqrt{1-4(mR)^2}\right)$. The heaviside theta function $\theta(x)$ indicates the fact that this complementary series contribution appears only when the scalar is light enough (in terms of the mass, when $m^2<\frac{3}{16R^2}$). The Dirac delta $\delta(x)$ implies the contributions are isolated states rather than a continuum. This seems to be a very general fact, there is no example in the literature where the complementary series of irreps appear as a continuum of states. 

It can be checked that these densities are positive\footnote{The density $ \varrho^{\mathcal{D}_2}_\Theta$ is positive for all $\lambda_\phi\in\mathbb{R}\cup i(-\frac{1}{2},\frac{1}{2})$. The complementary series density $\varrho^{\mathcal{C}}_{\Theta}$ is positive on the support of the Heaviside theta function in (\ref{eq:spectralTT}) after applying the Dirac delta}, and that evaluating the integrals in (\ref{eq:spectralTT}) numerically one can reproduce the analytic expression of $\langle T^{AB}T^{CD}\rangle$ from the Wick contractions of (\ref{eq:Tmunuboson}).

If one considers a massless and compact scalar from the start, as in \cite{Chakraborty:2023eoq}, then $G_\Theta(\sigma)=\varrho_\Theta^{\mathcal{P}}(\Delta)=\varrho^{\mathcal{C}}_{\Theta}(\Delta)=0$ and $\varrho^{\mathcal{D}_2}_{\hat T}=\frac{1}{12\pi R^4}$, giving
\begin{equation}
    \langle T^{AB}(Y_1)T^{CD}(Y_2)\rangle=\frac{1}{12\pi R^4}\Pi_1^{AB}\Pi^{CD}_2G_{\Delta=2}(\sigma)\,,
\end{equation}
meaning that in the massless case the stress tensor precisely creates states in the $\Delta=2$ irrep in the discrete series. This makes sense, given that this theory is conformally invariant and the stress tensor in a CFT is a spin 2 primary with $\Delta=2$.

\begin{figure}
\centering
\includegraphics[scale=0.7]{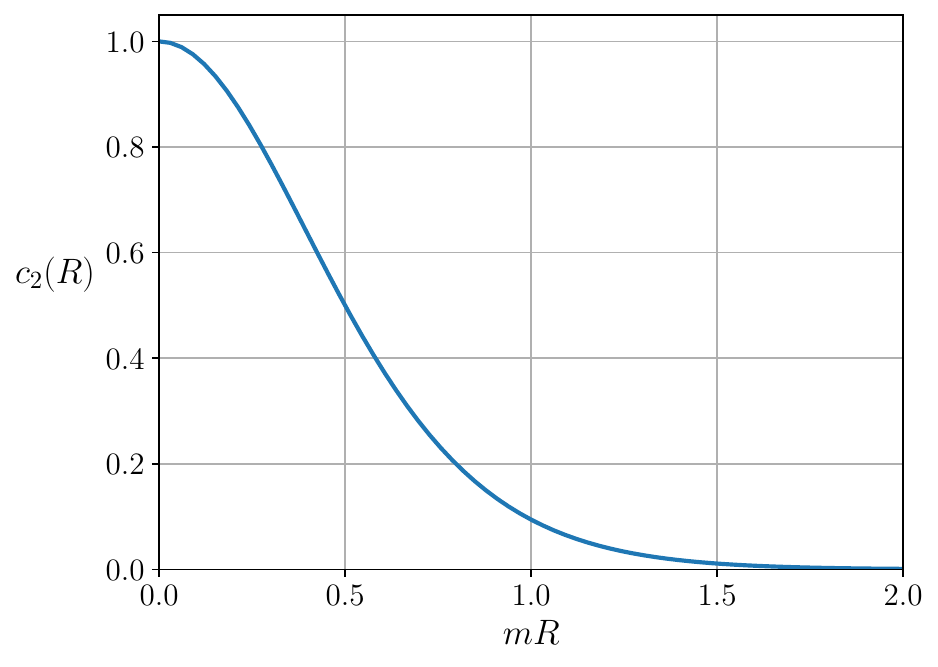}
\caption{Plot of the second c-function in the free massive scalar case, for which we derived an analytic expression, eq. (\ref{eq:c2boson}). It interpolates between $c^{\text{UV}}=1$, the CFT of the free massless scalar, and $c^{\text{IR}}=0$, the empty theory.}
\label{fig:boson}
\end{figure}

Now that we have all the spectral densities, we can check the individual formulas for $c^{\text{UV}}$ (\ref{eq:cuvsum}) and $c_1(R)$ (\ref{eq:crsum}). We find once again that $c^{\text{UV}}=1$ and that $c_1(R)=0$ for all $R$, due to the IR issues of the massless scalar theory in de Sitter. We can also compute the second c-function $c_2(R)$ from its definition (\ref{eq:defc2r}), and we obtain explicitly
\begin{equation}
    c_2(R)=4\pi m^2R^2\sqrt{m^2R^2-\frac{1}{4}}\text{csch}\left(2\pi\sqrt{m^2R^2-\frac{1}{4}}\right)\,.
    \label{eq:c2boson}
\end{equation}
We plot it in figure \ref{fig:boson}, and we observe that it indeed is a monotonic function which interpolates between $c^{\text{UV}}=1$ and $c^{\text{IR}}=0$. 
We also check that the sum rule (\ref{eq:newsumrule}) returns the same function, testing the fact that it works even when the complementary series contributes to the spectral decomposition of $G_\Theta(\sigma)$.

\begin{figure}
\centering
\includegraphics[scale=1.2]{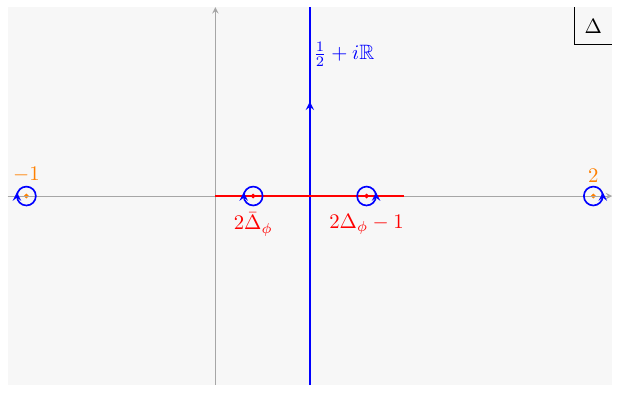}
\caption{In blue, the contour of integration $\gamma$ in (\ref{eq:contoursstress}). A vertical line runs over the principal series and circles surround the poles corresponding to a discrete series and a complementary series UIR contributing to the spectral decomposition of the stress tensor. Because of shadow symmetry, the residues on a pole at $\Delta$ and $1-\Delta$ are equal and opposite in sign. Here, we represented the case where the massive boson is in the complementary series and has $\Delta_\phi>3/4$.}
\label{fig:contours}
\end{figure}

Finally, notice that in this special case of the free massive boson, the full spectral decomposition of the stress tensor can be expressed as one spectral integral with a modified contour. That is because there exists a regime of the parameters of the theory, namely in $d>1$ and $\Delta_\phi\in\frac{d}{2}+i\mathbb{R}\cup(\frac{d}{4},\frac{3d}{4})$, for which only the principal series contributes. We show this in Appendix \ref{subsec:detailsboson}. Then, since the two-point function of the stress tensor is an analytic function of $d$ and $\Delta_\phi$, the only thing that can happen is that poles in the principal series spectral density cross the integration contour and lead to extra contributions to the K\"allén-Lehmann decomposition as we continue in the mass of the scalar and in the dimensions. These poles can be accounted for by modifying the contour of integration, leading to the following decomposition in two dimensions
\begin{equation}
    \langle T^{AB}(Y_1)T^{CD}(Y_2)\rangle=\int_{\gamma}\frac{d\Delta}{2\pi i}\frac{\varrho^{\mathcal{P}}_\Theta(\Delta)}{(\Delta+1)^2(\bar\Delta+1)^2}\Pi^{AB}_1\Pi^{CD}_2G_\Delta(\sigma)\,,
    \label{eq:contoursstress}
\end{equation}
with the contour $\gamma$ shown in blue in figure \ref{fig:contours}.
\subsection{Free massive fermion}
\label{sec:freefermion}
As a second example, consider the theory of a free Majorana fermion in two dimensions
\begin{equation}
    S=-\frac{1}{2}\int d^2x\sqrt{g}\bar\Psi\left(\slashed{\nabla}+m\right)\Psi\,.
    \label{eq:actionfermion}
\end{equation}
This can be seen as a specific field parametrization of two-dimensional Ising field theory on the sphere above the critical temperature and with zero magnetic field. The Ising CFT is reached when $m=0$, and it notoriously has $c^{\text{UV}}=\frac{1}{2}$. The mass term $m\bar\Psi\Psi$ acts as a deforming operator which triggers a flow to the trivial empty theory in the IR, which has $c^{\text{IR}}=0$. 

\begin{figure}
\centering
\includegraphics[scale=0.8]{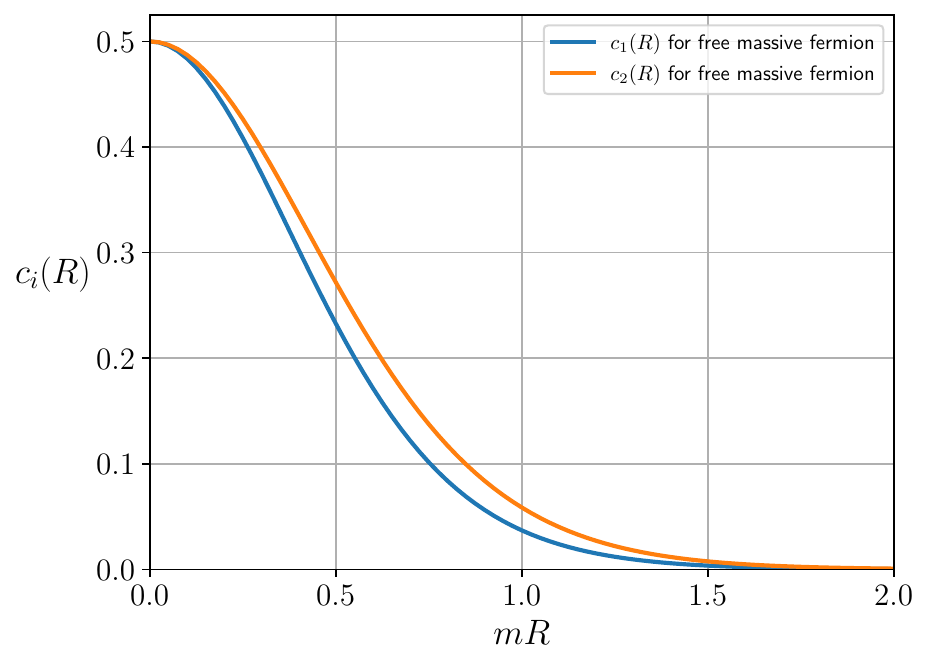}
\caption{Numerical plots of $c_1(R)$ and $c_2(R)$ for the free massive fermion, obtained by using (\ref{eq:sumrule2}) and (\ref{eq:newsumrule}) with the two-point function of the trace of the stress tensor in this theory (\ref{eq:thetafermion}), and knowing that $c^{\text{UV}}=1/2$. They interpolate between the critical Ising model in the UV and the empty theory in the IR.}
\label{fig:fermions}
\end{figure}

We leave many details to Appendix \ref{subsec:detailsfermion}. The canonically normalized two-point function is \cite{Schaub:2023scu,Pethybridge:2021rwf}
\begin{equation}
    \langle\Psi(x_1)\bar\Psi(x_2)\rangle=\frac{1}{\sqrt{\eta_1\eta_2}}\begin{pmatrix}
        i[(\eta_1+\eta_2)+(y_1-y_2)]G^-_m(\sigma) & [(\eta_1-\eta_2)+(y_1-y_2)]G^+_m(\sigma)\\
        [(y_1-y_2)-(\eta_1-\eta_2)]G^+_m(\sigma) & i[(\eta_1+\eta_2)-(y_1-y_2)]G^-_m(\sigma)
    \end{pmatrix}\,,
    \label{eq:freefermion}
\end{equation}
where
    \begin{equation}
\begin{aligned}
    G^+_m(\sigma)&\equiv\frac{1}{8}m\ \text{csch}(\pi mR)\ _2F_1\left(1-imR,1+imR,1,\frac{1+\sigma}{2}\right)\,,\\
    G^-_m(\sigma)&\equiv-\frac{i}{8}m^2R\ \text{csch}(\pi mR)\ _2F_1\left(1-imR,1+imR,2,\frac{1+\sigma}{2}\right)\,,
    \label{eq:gplusgminus}
\end{aligned}
\end{equation}
and we are working in flat slicing coordinates $ds^2=R^2\frac{-\mathrm{d}\eta^2+\mathrm{d}y^2}{\eta^2}$ and $x^\mu=(\eta,y)\,.$ The two-point invariant then takes the form
\begin{equation}
    \sigma=\frac{\eta_1^2+\eta_2^2-(y_1-y_2)^2}{2\eta_1\eta_2}\,.
\end{equation}
In Appendix \ref{subsec:detailsfermion} we show that, in the flat space limit, (\ref{eq:freefermion}) reduces to the canonically normalized two-point function of a free fermion in two-dimensional flat space. The symmetric and conserved stress tensor for this theory is \cite{Freedman:2012zz}
\begin{equation}
    T_{\mu\nu}=\frac{1}{8}\bar\Psi\left(\Gamma_\mu{\overset{\leftrightarrow}{\nabla}}_\nu+\Gamma_\nu{\overset{\leftrightarrow}{\nabla}}_\mu\right)\Psi\,,
\end{equation}
where $A{\overset{\leftrightarrow}{\nabla}}_\mu B\equiv A\left(\nabla_\mu B\right)-\left(\nabla_\mu A\right)B$, and $\Gamma_\mu$ are the Dirac gamma matrices in de Sitter, related to the flat space gamma matrices through the zweibein (see \ref{subsec:detailsfermion} for an explanation). Using the equations of motion, the trace of the stress tensor reduces to
\begin{equation}
    \Theta=-\frac{m}{2}\bar\Psi\Psi\,,
\end{equation}
with two-point function
\begin{equation}
    G_\Theta(\sigma)=\langle\Theta(x_1)\Theta(x_2)\rangle=2m^2\left((1-\sigma)\left(G_m^+(\sigma)\right)^2+(1+\sigma)\left(G_m^-(\sigma)\right)^2\right)\,.
    \label{eq:thetafermion}
\end{equation}
Applying our formula (\ref{eq:sumrule2}) we numerically verify that
\begin{equation}
    c^{\text{UV}}-\lim_{R\to\infty}c_1(R)=\frac{1}{2}\,.
\end{equation}
In contrast to the free boson case, it can be verified that in this theory
\begin{equation}
    \lim_{R\to0}G_\Theta(\sigma)=0\,,
\end{equation}
implying both (\ref{cuv10}) and (\ref{eq:criterc2}) are true, so that both c-functions will interpolate between $c^{\text{UV}}$ and $c^{\text{IR}}$. Using that $c^{\text{UV}}=\frac{1}{2}$, we show a numerical plot of $c_1(R)$ in figure \ref{fig:fermions}. It is a monotonically decreasing function of the radius.

To compute $c_2(R)$ for this theory, we start from the sum rule (\ref{eq:newsumrule}), which requires computing the integral of $r_2(\sigma)G_\Theta(\sigma)$ over the contour $\mathcal{C}_k$ shown in figure \ref{fig:keyhole}. We notice that (\ref{eq:thetafermion}) has a simple pole at $\sigma=1$ and a branch cut at $\sigma\in[1,\infty)$. The sum rule thus becomes
\begin{equation}
    c^{\text{UV}}-c_2(R)=\int_1^\infty d\sigma\ r_2(\sigma)\text{Disc}\left[R^4G_\Theta(\sigma)\right]-2\pi i\ \underset{\sigma=1}{\text{Res}}\left[r_2(\sigma)R^4G_\Theta(\sigma)\right]
\end{equation}
The discontinuity can be computed analytically using (\ref{eq:discontinuities}). The residue is simply
\begin{equation}
    2\pi i\ \underset{\sigma=1}{\text{Res}}\left[r_2(\sigma)R^4G_\Theta(\sigma)\right]=m^2R^2\left(2-\frac{\pi^2}{3}\right)\,.
\end{equation}
We evaluate the remaining integral numerically and plot the function $c_2(R)$ in figure \ref{fig:fermions}. It is also a monotonic function of $R$, and it satisfies the condition (\ref{eq:c2gc1}).
\subsection{A comment on the massless Schwinger model}
The massless Schwinger model is an integrable QFT in dS$_2$ \cite{Anninos:2024fty,Jayewardena:1988td,Shore:1978hj}. Here we show that its associated functions $c_1$ and $c_2$ are precisely the same as in the free massive scalar theory, indicating that the two theories are related by a field redefinition as in flat space. The massless Schwinger model is defined through the following action
\begin{equation}
    S=\int d^2x\sqrt{g}\left[\bar\Psi\left(\slashed{\nabla}+i\slashed{A}\right)\Psi+\frac{1}{4q^2}F^{\mu\nu}F_{\mu\nu}\right]
    \label{eq:actionschwinger}
\end{equation}
where $\Psi$ is a Dirac spinor, $A_{\mu}$ is a compact $U(1)$ gauge field with field strength $F_{\mu\nu}=\partial_\mu A_\nu-\partial_\nu A_\mu$ and $q$ is the gauge coupling, which has mass dimensions 1. The trace of the stress tensor (\ref{eq:defTmunu}) is, on-shell,
\begin{equation}
    \Theta=\frac{1}{2q^2}F^{\mu\nu}F_{\mu\nu}
\end{equation}
The two-point function of $F$ in the Bunch-Davies vacuum is written explicitly in \cite{Anninos:2024fty}, and it has the precise form of the two-point function of a free massive boson (\ref{eq:freeprop}):
\begin{equation}
    \frac{1}{\sqrt{g(x_1)g(x_2)}}\langle F_{01}(x_1)F_{01}(x_2)\rangle=-\frac{q^4}{\pi}G_{\Delta_q}(\sigma)\,,
\end{equation}
where in this case $\Delta_q(1-\Delta_q)=\frac{1}{\pi}q^2R^2$. Since in two dimensions this is the only degree of freedom of the field strength, this implies
\begin{equation}
    G_\Theta(\sigma)=2\left(\frac{q^2}{\pi}\right)^2\left(G_{\Delta_q}(\sigma)\right)^2\,.
\end{equation}
This is exactly the same two-point function as in (\ref{eq:thetaboson}). Since $c_1$ and $c_2$ can be derived through sum rules (\ref{eq:sumrule2}) and (\ref{eq:newsumrule}) which only depend on the trace of the stress tensor, they are precisely the same as for the free massive boson theory, up to the mapping $m^2\leftrightarrow\frac{q^2}{\pi}$. This is not unexpected: it is well known in flat space that the massless Schwinger model can be mapped through a field redefinition to the free massive boson theory, precisely with $m^2\leftrightarrow\frac{q^2}{\pi}$ \cite{Lowenstein:1971fc,Iso:1988zj}. This equality of the c-functions hints to the fact that carefully bosonizing the action (\ref{eq:actionschwinger}) should lead to the free massive scalar theory in dS$_2$ as well.
\section{Discussion}
\label{sec:discussion}
In this work we have studied RG flows in unitary QFTs in dS$_2$ and $S^2$. We have introduced two functions of the radius which interpolate between the central charges of the CFTs that live at the fixed points of any RG flow. One is defined through certain components of the two-point function of the stress tensor at antipodal separation (\ref{eq:defcR}), while the other is the spectral weight of the traceless part of the stress tensor in the $\Delta=2$ irrep (\ref{eq:defc2r}). The fact that this spectral weight has to interpolate between the two central charges implies that it needs to be non-zero for any QFT, or in other words the stress tensor has to always couple to discrete series $\Delta=2$ states. We have verified our formulas in the examples of the theories of a free massive boson, a free massive fermion and the Schwinger model. We showed that $c_2$ is monotonically decreasing in every case, while $c_1$ is monotonically decreasing in the free fermion flow and it is zero for all radii in the free boson case where the massless regime is ill-defined due to IR divergences. We found that the massless Schwinger model has the same $c_1$ and $c_2$ as the free boson theory. As an intermediate step, we worked out the details of how the conservation of the stress tensor simplifies its spectral decomposition greatly. We argue that, in general, the sphere and de Sitter can be interesting background geometries to study QFT since the radius acts as a tunable parameter which, while not breaking any symmetry, can be used to follow the RG flow and reveal new facts about QFTs of interest which may be inaccessible in flat space. Moreover, the existence and behavior of $c_1$ and $c_2$ are new rigorous constraints that any unitary QFT in dS$_2$ must satisfy.

There are some open questions which would be interesting to explore in the future:
\begin{itemize}
	\item The c-functions we have introduced interpolate between $c^{\text{UV}}$ and $c^{\text{IR}}$. We also showed that $c^{\text{UV}}\geq c_i(R)$ for both, implying in particular Zamolodchikov's c-theorem $c^{\text{UV}}\geq c^{\text{IR}}$. In the two examples we studied, we also verified that they are monotonic for intermediate radii. It would be interesting to establish whether the monotonicity is true for all QFTs with a general proof or a counterexample.
    \item The examples in which we could test our formulas were gapped theories. In the future, we hope to test them in flows which have $c^{\text{IR}}\neq 0$, such as between minimal models in de Sitter. 
    \item Can a similar approach to the one utilized in this work be adapted to the problem of finding RG-monotonic functions constructed from the stress tensor two-point function in higher dimensions? Analogously to what happens in AdS \cite{Meineri:2023mps}, the simple generalization of the differential equation (\ref{eq:diffeq}) to higher dimensions is not enough to extract the trace anomalies, so a more sophisticated approach is required.
    \item The results of this paper can be thought of as a new set of constraints that any unitary QFT in $S^2/$dS$_2$ needs to satisfy. Some of them are in the form of positive sum rules on two-point functions of the stress tensor which relate IR and UV data. Combining these with constraints on higher-point functions one may be able to set up numerical bootstrap problems in de Sitter, as suggested in \cite{Hogervorst:2021uvp,Penedones:2023uqc,Loparco:2023rug,Loparco:2023akg,Kravchuk:2021akc,DiPietro:2021sjt,Gesteau:2023brw}.
    \item Are there any RG flows of interest for which the approach presented in this paper is more efficient than the well known flat space techniques? It would be interesting to understand whether there are computational advantages that come, for example, from the fact that one of the c-functions we propose is only dependent on the $\Delta=2$ contribution to the spectral decomposition of the traceless part of the stress tensor, which has the form of a CFT two-point function of a spin 2 primary on the sphere. 
\end{itemize}
\section*{Acknowledgements}
I am grateful to Dionysios Anninos, Tarek Anous, Victor Gorbenko, Grégoire Mathys, Joao Penedones, Jiaxin Qiao, Veronica Sacchi, Vladimir Schaub, Zimo Sun, Kamran Salehi Vaziri and Antoine Vuignier for insightful discussions. I am especially grateful to Joao Penedones for his consistent and supportive supervision, Vladimir Schaub for useful discussions about fermions in de Sitter and Grégoire Mathys for conversations on c-theorems and sum rules in flat space. Finally, I thank Dionysios Anninos and Tarek Anous for hospitality at King's College and Queen Mary University in December 2023, a visit from which I learned a lot. I am supported by the Simons Foundation
grant 488649 (Simons Collaboration on the Nonperturbative Bootstrap) and the Swiss
National Science Foundation through the project 200020\_197160 and through the National
Centre of Competence in Research SwissMAP.
\appendix
\section{Details on the position space sum rules}
\label{sec:appproof}
In this appendix we report some details concerning section (\ref{sec:proofposition}) of the main text.
\subsection{Conservation equations and solutions of (\ref{eq:diffeq})}
Here we report the three linearly independent constraints we get from imposing the conservation of the stress tensor on the functions $T_i(\sigma)$. 
\small
\begin{align}
        \frac{\sigma^2-1}{\sigma^3}\Big[&-(d+4)\sigma^4(\sigma^2-1)T_1(\sigma)+(d^2+3d+4)\sigma^4T_2(\sigma)+4(\sigma+(d+1)\sigma^3)T_3(\sigma)\nonumber\\
        &+(4+2d\sigma^2)T_5(\sigma)-\sigma^3(\sigma^2-1)^2T'_1(\sigma)+(d+2)\sigma^3(\sigma^2-1)T_2'(\sigma)\nonumber\\
        &+4\sigma^2(\sigma^2-1)T_3'(\sigma)-(d+1)\sigma^3T_4'(\sigma)-2\sigma T_5'(\sigma)\Big]=0\,,\nonumber\\
        \frac{(\sigma^2-1)^2}{\sigma^3}\Big[&(d+4)\sigma^4(\sigma^2-1)T_1(\sigma)-(d+4)\sigma^4T_2(\sigma)-2\sigma(2+(d+2)\sigma^2)T_3(\sigma)\nonumber\\
        &-4T_5(\sigma)+\sigma^3(\sigma^2-1)^2T'_1(\sigma)-2\sigma^3(\sigma^2-1)T'_2(\sigma)-4\sigma^2(\sigma^2-1)T'_3(\sigma) \label{eq:conservationeqs}\\
        &+\sigma^3T'_4(\sigma)+2\sigma T'_5(\sigma)\Big]=0\,,\nonumber\\
        \frac{\sigma^2-1}{\sigma^4}\Big[&(d+4)\sigma^4(\sigma^2-1)T_1(\sigma)-2(d+2)\sigma^4T_2(\sigma)-\sigma(4+2(d+2)\sigma^2+d(d+3)\sigma^4)T_3(\sigma)\nonumber\\
        &-(4+d\sigma^2)T_5(\sigma)+\sigma^3(\sigma^2-1)^2T_1'(\sigma)-2\sigma^3(\sigma^2-1)T_2'(\sigma)\nonumber\\
        &-\sigma^2(\sigma^2-1)(4+d\sigma^2)T_3'(\sigma)+\sigma^3T_4'(\sigma)+\sigma(2+d\sigma^2)T_5'(\sigma)\Big]=0\,.     \nonumber
\end{align}
\normalsize
They are obtained, respectively, by acting on $\nabla^A\langle T_{AB}(Y_1)T_{CD}(Y_2)\rangle=0$ with the projectors
\begin{equation}
    V_1^BG_2^{CD}\,, \qquad V_1^BV_2^CV_2^D\,, \qquad G_{12}^{BC}V_2^D\,,
\end{equation}
where the explicit form of these objects is in equation (\ref{eq:GVblobs}).
In two dimensions, these are differential equations for the four $\mathcal{T}_i$ functions 
\small
\begin{equation}
\begin{aligned}
    (2+\sigma^2)\mathcal{T}_1(\sigma)+\left(\sigma^2+\frac{2}{\sigma^2-1}\right)\mathcal{T}_2(\sigma)-(\sigma^2+2)\mathcal{T}_3(\sigma)+\left(2+\sigma^2\right)\mathcal{T}_4(\sigma)&\\
    +\sigma\left(\sigma^2-1\right)\mathcal{T}_1'(\sigma)+\sigma\left(1+2\sigma^2\right)\mathcal{T}'_2(\sigma)+\sigma\left(1-2\sigma^2\right)\mathcal{T}_3'(\sigma)-\sigma\mathcal{T}_4'(\sigma)&=0\,,\\
    -(\sigma^2+2)\mathcal{T}_1(\sigma)+\frac{\sigma^2-2}{\sigma^2-1}\mathcal{T}_2(\sigma)+2\mathcal{T}_3(\sigma)-2\mathcal{T}_4(\sigma)&\\
    +\sigma(1-\sigma^2)\mathcal{T}_1'(\sigma)-\sigma(\sigma^2+1)\mathcal{T}_2'(\sigma)+\sigma(\sigma^2-1)\mathcal{T}_3'(\sigma)+\sigma\mathcal{T}_4'(\sigma)&=0\,,\\
  -(2+\sigma^2)\mathcal{T}_1(\sigma)+\frac{\sigma^4-\sigma^2+2}{1-\sigma^2}\mathcal{T}_2(\sigma)+\frac{1}{2}(\sigma^2+4)\mathcal{T}_3(\sigma)-\frac{1}{2}(4+\sigma^2)\mathcal{T}_4(\sigma)&\\
  +\sigma(1-\sigma^2)\mathcal{T}_1'(\sigma)-\sigma(\sigma^2+1)\mathcal{T}_2'(\sigma)+\frac{\sigma}{2}(\sigma^2-2)\mathcal{T}_3'(\sigma)+\frac{\sigma}{2}(2+\sigma^2)\mathcal{T}_4'(\sigma)&=0\,.
    \label{eq:conservetildeT}
\end{aligned}
\end{equation}
\normalsize
And here we report the kernels and functions $C$ and $q_i$ that solve (\ref{eq:diffeq}), parametrized by three real numbers $c_1,\ c_2,\ c_3\,.$
\small
\begin{equation}
    \begin{aligned}
       r(\sigma)=&\frac{1}{2}\left[2c_3\sigma-2c_2+\sigma(c_1-c_2)\log(1-\sigma)+\sigma(c_1+c_2)\log(1+\sigma)\right]\,,\\
       C(\sigma)=&\frac{1}{\sigma^2}\Big[-\left(1-\sigma ^2\right)^2(2(c_3-c_1)+(c_1-c_2)\log(1-\sigma)+(c_1+c_2)\log(1+\sigma))\mathcal{T}_1(\sigma)\\
       &\qquad\qquad+(2((\sigma^4-2\sigma^2-1)c_1+c_3+\sigma^2(c_3+2\sigma(c_2-\sigma c_3)))\\
       &\qquad\qquad+(1-\sigma^2)(2\sigma^2+1)(2\tanh^{-1}(\sigma) c_2+c_1\log(1-\sigma^2)))\mathcal{T}_2(\sigma)\\
       &\qquad\qquad+(-2(\sigma^4-5\sigma^2+1)c_1-\sigma(1+5\sigma^2)c_2+2(2\sigma^4-3\sigma^2+1)c_3\\
       &\qquad\qquad+(2\sigma^4-3\sigma^2+1)(2\tanh^{-1}(\sigma) c_2+c_1\log(1-\sigma^2)))\mathcal{T}_3(\sigma)\\
       &\qquad\qquad+((2-4\sigma^2)c_1+\sigma(\sigma^2+1)c_2+2(\sigma^2-1)c_3\\
       &\qquad\qquad+(1-\sigma^2)((c_2-c_1)\log(1-\sigma)-(c_1+c_2)\log(\sigma+1)))\mathcal{T}_4(\sigma)\Big]\\
       q_1(\sigma)&=\frac{1}{2\sigma^3}\left[4\sigma c_2-4c_1+(1-\sigma^2)\left(2c_3+(c_1-c_2)\log(1-\sigma)+(c_1+c_2)\log(\sigma+1)\right)\right]\\
       q_2(\sigma)&=\frac{1-\sigma^2}{\sigma^4}(\sigma c_1-c_2)\\
       q_3(\sigma)&=\frac{(1+\sigma^2)c_2-2\sigma c_1}{\sigma^4}
    \label{eq:threesols}
    \end{aligned}
\end{equation}
\normalsize
The $g_i(\sigma)$ functions can be read off from the expression for $C(\sigma)$ through their definition (\ref{eq:Cansatz}). 
The particular solution that leads to (\ref{eq:sumrule2}) is $c_1=c_2=-8\pi^2$ and $c_3=8\pi^2(\log(2)-1)$. 
\subsection{Coincident point limit of $T_i$}
\label{subsec:coincident}
Here we derive the coincident point limit of the $T_i(\sigma)$ functions in (\ref{eq:twopointstructs}). We use the following local coordinate system for de Sitter
\begin{equation}
    Y^0=\frac{1-e^{-2t}+\mathbf{x}^2}{2e^{-t}}\,, \qquad Y^i=x^ie^t\,, \qquad Y^{d+1}=\frac{-1-e^{-2t}+\mathbf{x}^2}{2e^{-t}}\,,
\end{equation}
where $\mathbf{x}\in\mathbb{R}^d$ with $i=1,\ldots,d$ and we keep $R=1$. In this coordinate system, the metric is 
\begin{equation}
    ds^2=-\mathrm{d}t^2+e^{2t}\mathrm{d}\mathbf{x}^2\,.
\end{equation}
The two-point invariant reads
\begin{equation}
    \sigma=\frac{1}{2}e^{-(t_1+t_2)}\left(e^{2t_1}+e^{2t_2}+2 e^{2(t_1+t_2)}\mathbf{x}_1\cdot\mathbf{x}_2-e^{2(t_1+t_2)}(\mathbf{x}_1^2+\mathbf{x}_2^2)\right)\,.
\end{equation}
Then, by using (\ref{eq:embtoloc}) we can compute the behavior of the tensor structures (\ref{eq:tensorstructs}) near coincident points in these local coordinates, where $x^{\mu}=(t,\mathbf{x})$.
\begin{equation}
\begin{aligned}
    \mathbb{T}_1^{\mu\nu\rho\sigma}&\approx x^\mu x^\nu x^\rho x^\sigma\,, \qquad \mathbb{T}_2^{\mu\nu\rho\sigma}\approx \eta^{\mu\nu}x^\rho x^\sigma+x^\mu x^\nu\eta_{\rho\sigma}-x^\mu x^\nu x^\rho x^\sigma\,,\\
    \mathbb{T}_3^{\mu\nu\rho\sigma}&\approx \eta^{\nu\sigma}x^\mu x^\rho+\eta^{\mu\sigma}x^\nu x^\rho+\eta^{\nu\rho}x^{\mu}x^\sigma+\eta^{\mu\rho}x^\nu x^\sigma\,,\\
    \mathbb{T}_4^{\mu\nu\rho\sigma}&\approx \eta^{\rho\sigma}\eta^{\mu\nu}-\eta^{\rho\sigma}x^{\mu}x^{\nu}\,, \qquad \mathbb{T}_5^{\mu\nu\rho\sigma}\approx \eta^{\mu\sigma}\eta^{\nu\rho}+\eta^{\mu\rho}\eta^{\nu\sigma}\,.
\end{aligned}
\end{equation}
This means the coincident point limit of our parametrization of the stress tensor (\ref{eq:twopointstructs}) is
\begin{equation}
\begin{aligned}
    \langle T^{\mu\nu}(x)T^{\rho\sigma}(0)\rangle\approx &\ x^\mu x^\nu x^\rho x^\sigma T_1(x)+(\eta^{\mu\nu} x^\rho x^\sigma+\eta^{\rho\sigma}x^\mu x^\nu)T_2(x)\\
    &+(\eta^{\nu\sigma}x^\mu x^\rho+\eta^{\mu\sigma}x^\nu x^\rho+\eta^{\nu\rho}x^\mu x^\sigma+\eta^{\mu\rho}x^\nu x^\sigma)T_3(x)\\
    &+\eta^{\mu\nu}\eta^{\rho\sigma}T_4(x)+(\eta^{\mu\sigma}\eta^{\nu\rho}+\eta^{\mu\rho}\eta^{\nu\sigma})T_5(x)\,.
    \label{eq:coincidentT}
\end{aligned}
\end{equation}
We need to match with the well known flat space CFT two point function of the stress tensor (\ref{eq:flatCFT}). We can reshuffle the expression (\ref{eq:flatCFT}) in order to expand it in the same tensor structures
\begin{align}
    \langle T^{\mu\nu}(x)T^{\rho\sigma}(0)\rangle^{\text{CFT}}_{\mathbb{M}}=&\frac{4c_T}{x^{2d+6}}x^\mu x^\nu x^\rho x^\sigma-\frac{c_T}{x^{2d+4}}(\eta^{\nu\sigma}x^\mu x^\rho+\eta^{\mu\sigma}x^\nu x^\rho+\eta^{\nu\rho}x^\mu x^\sigma+\eta^{\mu\rho}x^\nu x^\sigma)\nonumber\\
    &-\frac{c_T}{(d+1)x^{2d+2}}\eta^{\mu\nu}\eta^{\rho\sigma}+\frac{c_T}{2x^{2d+2}}(\eta^{\mu\sigma}\eta^{\nu\rho}+\eta^{\mu\rho}\eta^{\nu\sigma})
\end{align}
By matching with our $T_i$ functions, we find the constraints mentioned in the main text
\begin{equation}
\begin{aligned}
    &T_1\approx\frac{4c_T}{x^{2d+6}}\,, \qquad T_2\sim o(x^{-2d-2})\,,\qquad T_3\approx -\frac{c_T}{x^{2d+4}}\,,\\
    &T_4\approx-\frac{c_T}{d+1}\frac{1}{x^{2d+2}}\,,\qquad T_5\approx\frac{c_T}{2}\frac{1}{x^{2d+2}}\,.
\end{aligned}
\end{equation}
\section{Details on the spectral decomposition of the stress tensor}
\label{sec:appspectr}
In this section we provide extra details regarding the spectral decomposition of the stress tensor. First, we prove the relations (\ref{eq:spectralrelations}) used extensively in the main text. Then, we relate the $T_i(\sigma)$ and the $\mathcal{T}_i(\sigma)$ defined in (\ref{eq:twopointstructs}) and (\ref{eq:TtotildeT}) to integrals over spectral densities. This in turn allows us to prove that these functions are always finite at $\sigma=-1$, a fact that is crucial in deriving the form of $c_1(R)$ in (\ref{eq:defcR}). 
\subsection{General relations among the spectral densities}
\label{subsec:spectralrelations}
Here we prove the relations (\ref{eq:spectralrelations}) used in the main text. We start from the most general spectral decomposition of a spin 2 symmetric tensor in dS$_{d+1}$/$S^{d+1}$, for which we already explained the notation in Section \ref{sec:stresstensor}
 \begin{align}
         \langle T^{AB}(Y_1)T^{CD}(Y_2)\rangle=2\pi\int_{\frac{d}{2}+i\mathbb{R}}\frac{d\Delta}{2\pi i}\Bigg[&\sum_{\ell=0}^2\varrho_{\hat T,\ell}^{\mathcal{P}}(\Delta)G^{AB,CD}_{\Delta,\ell}(Y_1,Y_2)\nonumber\\
         &+\varrho^{\mathcal{P}}_{\hat T\Theta}(\Delta)\left(\frac{G_2^{CD}}{d+1}\hat\Pi_1^{AB}G_{\Delta}(\sigma)+\frac{G_1^{AB}}{d+1}\hat\Pi_2^{CD}G_{\Delta}(\sigma)\right)\nonumber\\
    &+\varrho^{\mathcal{P}}_\Theta(\Delta)\frac{G_1^{AB}G_2^{CD}}{(d+1)^2}G_\Delta(\sigma)\Bigg]+\text{other UIRs}\,,
    \end{align}
where the other UIRs can be complementary series, which has the same analytic expression as the principal series but is integrated over $\Delta\in(0,1)$, exceptional series type I and exceptional series type II \cite{Sun:2021thf,Loparco:2023rug}.
We impose conservation, so from now on we consider the equation
\begin{equation}
\nabla_A\langle T^{AB}(Y_1)T^{CD}(Y_2)\rangle=0\,. 
\label{eq:conservation}
\end{equation}
Various manipulations of this equation will lead to the spectral relations (\ref{eq:spectralrelations}). The complementary series contributions have the same functional form as the principal series ones, so every relation we are going to find is valid also for the spectral densities on the complementary series. We thus drop the superscript $\mathcal{P}$ and consider a generic contour for the integral over $\Delta$. We now start to consider the consequences of (\ref{eq:conservation}).

First of all, the divergence kills the $\ell=2$ term in the sum, which is automatically and independently conserved by definition. The first nontrivial statement comes from taking a trace over the indices $C,D$
\begin{equation}
    \int d\Delta\Big[\varrho_{\hat T\Theta}(\Delta)\nabla_{1A}\hat\Pi^{AB}+\frac{\varrho_\Theta(\Delta)}{d+1}\nabla_{1A}G^{AB}_1\Big]G_\Delta(\sigma)=0 
    \label{eq:steprhothetat}
\end{equation}
Using the explicit expressions of the projectors (\ref{eq:pihatdef}), the induced metric and the covariant derivative (\ref{eq:metricderivs}), we find that (\ref{eq:steprhothetat}) implies
\begin{equation}
    \varrho_{\hat T\Theta}(\Delta)=\frac{\varrho_{\Theta}(\Delta)}{d(\Delta+1)(\bar\Delta+1)}\,.
    \label{eq:spectral2}
\end{equation}
Now we use this fact, and (\ref{eq:conservation}) becomes
\begin{equation}
\begin{aligned}
    \int d\Delta\Big[&\varrho_{\hat T,1}(\Delta)\nabla_{1A}G^{AB,CD}_{\Delta,1}(Y_1,Y_2)+\varrho_{\hat T,0}(\Delta)\nabla_{1A}\hat\Pi^{AB}_1\hat\Pi^{CD}_2G_\Delta(\sigma)\\
    &+\varrho_{\Theta}(\Delta)\Big(\frac{1}{d(\Delta+1)(\bar\Delta+1)}\Big(\nabla_{1A}\hat\Pi^{AB}_1\frac{G_2^{CD}}{d+1}+\nabla_{1A}\frac{G_1^{AB}}{d+1}\hat\Pi_2^{CD}\Big)\\
    &\qquad\qquad\qquad\qquad\qquad\qquad\qquad\qquad\qquad+\nabla_{1A}\frac{G^{AB}_1G^{CD}_2}{(d+1)^2}\Big)G_\Delta(\sigma)\Big]=0\,,
\end{aligned}
\end{equation}
where we used that by definition $G_{\Delta,0}^{AB,CD}=\hat\Pi_1^{AB}\hat\Pi_2^{CD}G_\Delta$. Now, carrying out all the necessary computations, we find the last two relations
\begin{equation}
    \varrho_{\hat T,0}(\Delta)=\frac{\varrho_\Theta(\Delta)}{d^2(\Delta+1)^2(\bar\Delta+1)^2}\,, \qquad \varrho_{\hat T,1}(\Delta)=0\,,
    \label{eq:spectral3}
\end{equation}
where we used the fact that the term proportional to $\varrho_{\hat T,0}$ turns out to have the same tensor structure as the one proportional to $\varrho_\Theta$, while $\varrho_{\hat T,1}$ has an independent tensor structure, and thus has to vanish on its own. This shows a fact that is well known in flat space: the stress tensor cannot interpolate between the vacuum and states carrying $SO(d)$ spin 1.

Using all of these relations, the expressions simplify greatly, and all the terms associated to $\varrho_\Theta$, $\varrho_{\hat T\Theta}$ and $\varrho_{\hat T,0}$ collapse into one single term, resulting in
\begin{equation}
\begin{aligned}
    \langle T^{AB}(Y_1)T^{CD}(Y_2)\rangle=&2\pi\int_{\frac{d}{2}+i\mathbb{R}}\frac{d\Delta}{2\pi i}\Big[\varrho^{\mathcal{P}}_{\hat T,2}(\Delta)G^{AB,CD}_{\Delta,2}(Y_1,Y_2)\\
    &+\frac{\varrho^{\mathcal{P}}_{\Theta}(\Delta)}{d^2(\Delta+1)^2(\bar\Delta+1)^2}\Pi_1^{AB}\Pi^{CD}_2G_\Delta(\sigma)\Big]\\
    &+\text{other UIRs}\,,
\end{aligned}
\end{equation}
with $\Pi^{AB}_i$ defined in (\ref{eq:definepi}).

In \textbf{two dimensions}, the argument is very similar. The most general decomposition of a spin 2 symmetric tensor is 
    \begin{align}
    \langle T^{AB}(Y_1)T^{CD}(Y_2)\rangle=&2\pi\int_{\frac{1}{2}+i\mathbb{R}}\frac{d\Delta}{2\pi i}\Bigg[ \varrho^{\mathcal{P}}_{\hat T,1}(\Delta)G_{\Delta,1}^{AB,CD}(Y_1,Y_1)+\varrho^{\mathcal{P}}_{\hat T,0}(\Delta)\hat\Pi^{AB}_1\hat\Pi^{CD}_2G_\Delta(\sigma)\nonumber\\
    &+\varrho_{\hat T\Theta}^{\mathcal{P}}(\Delta)\left(\hat\Pi_1^{AB}\frac{G_2^{CD}}{2}G_{\Delta}(\sigma)+\frac{G_1^{AB}}{2}\hat\Pi_2^{CD}G_{\Delta}(\sigma)\right)\nonumber\\
    &+\frac{G_1^{AB}G_2^{CD}}{4}\varrho^{\mathcal{P}}_\Theta(\Delta)G_\Delta(\sigma)\Bigg]+\text{complementary series}\\
    &+\varrho^{\mathcal{D}_1}_{\hat T}\hat\Pi_1^{AB}\hat\Pi_2^{CD}G_{\Delta=1}(\sigma)+\varrho^{\mathcal{D}_2}_{\hat T}\hat\Pi_1^{AB}\hat\Pi_2^{CD}G_{\Delta=2}(\sigma)\nonumber
\end{align}
where, group theoretically, the first two terms stand for the contributions from states carrying the two inequivalent chiralities of $SO(1,2)$ (see \cite{Loparco:2023rug} for an in-depth discussion on this), and of course states do not carry any spin in two dimensions, so there is no $\ell=2$ contribution. The only difference between the complementary and principal series contributions will be again the domain of integration, while the discrete series has been explicitly added. Contributions from this series of UIRs can only be traceless because the trace of the stress tensor is a scalar operator, and as such it cannot carry discrete series irreps. More in general, local operators with spin $J$ can only couple to discrete series states with $\Delta\leq J$ \cite{Loparco:2023akg,Loparco:2023rug}.

Now we impose conservation (\ref{eq:conservation}). For the principal and complementary series contributions, the computations are analogous to what was done in the previous paragraph, and so one obtains (\ref{eq:spectral2}) and (\ref{eq:spectral3}) with $d=1$. For the discrete series, something interesting happens: we have that
\begin{equation}
    \nabla_{1A}\hat\Pi_1^{AB}\hat\Pi_{2}^{CD}G_{\Delta=2}(\sigma)=0\,, \qquad \nabla_{1A}\hat\Pi_1^{AB}\hat\Pi_{2}^{CD}G_{\Delta=1}(\sigma)\neq0\,.
\end{equation}
This implies that necessarily $\varrho^{\mathcal{D}_1}_{\hat T}(\Delta)=0$, or in other words the stress tensor cannot interpolate between the vacuum and states in the $\Delta=1$ discrete series irrep. Instead, the $\Delta=2$ irrep is allowed and is conserved independently from the principal and complementary series contributions. Using all these facts together, we can write
\begin{equation}
\begin{aligned}
    \langle T^{AB}(Y_1)T^{CD}(Y_2)\rangle=&2\pi\int_{\frac{1}{2}+i\mathbb{R}}\frac{d\Delta}{2\pi i}\ \frac{\varrho^{\mathcal{P}}_{\Theta}(\Delta)}{(\Delta+1)^2(\bar\Delta+1)^2}\Pi_1^{AB}\Pi^{CD}_2G_\Delta(\sigma)\\
    &+\int_{0}^1d\Delta\ \frac{\varrho^{\mathcal{C}}_{\Theta}(\Delta)}{(\Delta+1)^2(\bar\Delta+1)^2}\Pi_1^{AB}\Pi^{CD}_2G_\Delta(\sigma)\\
    &+\varrho_{\hat T}^{\mathcal{D}_2}\Pi_1^{AB}\Pi_2^{CD}G_{\Delta=2}(\sigma)\,,
\end{aligned}
\end{equation}
which is the complete and general spectral decomposition of the stress tensor two-point function in $S^2/\text{dS}_2$. In Section \ref{sec:stresstensor} we have shown that in any unitary QFT $\varrho_{\hat T}^{\mathcal{D}_2}$ interpolates between $c^{\text{UV}}$ and $c^{\text{IR}}$. This implies that the $\Delta=2$ contribution to the spectral decomposition of the stress tensor is not just \emph{allowed} but rather it is \emph{necessary} in any unitary QFT in $S^2/\text{dS}_2$.
\subsection{How to compute the $T_i$ functions}
\label{subsec:Tifuncs}
Throughout this work, we have used two decompositions of the stress tensor, namely the ones in terms of tensor structures (\ref{eq:twopointstructs}) and the ones in terms of the spectral densities (\ref{eq:stresstensorspectral}), (\ref{eq:stresstensorspectral2d}). Here, we are going to show relations between the two, which are crucial in deriving formulas for $c^{\text{UV}}$ and $c_1(R)$ independently of their difference. The main idea is to look closely at the explicit expressions of the tensor structures $\mathbb{T}_i$ given in (\ref{eq:tensorstructs}). We notice that there are some combinations of the coordinates and the metric with specific indices that appear uniquely in each tensor structure. Specifically,
\begin{equation}
    \begin{aligned}
          \mathbb{T}_5^{ABCD}&\overset{!}{\supset}\{\eta^{AD}\eta^{BC},\eta^{AC}\eta^{BD}\}\,,\\
    \mathbb{T}_4^{ABCD}&\overset{!}{\supset}\{\eta^{AB}\eta^{CD}\}\,,\\
    \mathbb{T}_3^{ABCD}&\overset{!}{\supset}\{\sigma\eta^{BD}Y_1^AY_1^C,\sigma\eta^{AD}
    Y_1^BY_1^C\}\,,\\
    \mathbb{T}_2^{ABCD}&\overset{!}{\supset}\{\eta^{AB}Y_1^CY_1^D,-\sigma\eta^{CD}Y_1^BY_2^A,-\sigma\eta^{CD}Y_1^AY_2^B,\sigma\eta^{CD} Y_2^AY_2^B\}\,.
    \label{eq:uniquestructs}
    \end{aligned}
\end{equation}
whre with the symbol $\overset{!}{\supset}$ we mean the terms on the right hand side appear only in the tensor structure on the left hand side and not in the others.

Finding the coefficients of any of the terms on the right hand side thus uniquely identifies the $T_i$ component within a two-point function. For $T_1$, it is sufficient to subtract all other contributions. Here, we will do this in (\ref{eq:stresstensorspectral}) in order to find relations between the $T_i$ and integrals of spectral densities. We obtain
\begin{align}
    T_5(\sigma)&=2\pi\int_{\frac{d}{2}+i\mathbb{R}}\frac{d\Delta}{2\pi i}\left(\frac{1}{2}\varrho^{\mathcal{P}}_{\hat{T},2}(\Delta)\mathcal{G}_0(\sigma)+\varrho^{\mathcal{P}}_{\hat{T},0}(\Delta)G''_{\Delta}(\sigma)\right)+\ldots \nonumber\\
    T_4(\sigma)&=2\pi\int_{\frac{d}{2}+i\mathbb{R}}\frac{d\Delta}{2\pi i}\Big(\frac{\varrho^{\mathcal{P}}_{\hat{T},2}(\Delta)}{(d+1)^2}\left((1-\sigma^2)^2\mathcal{G}_2(\sigma)+\sigma(\sigma^2-1)\mathcal{G}_1(\sigma)+(\sigma^2-d-2)\mathcal{G}_0(\sigma)\right)\nonumber\\
    &\qquad+\varrho^{\mathcal{P}}_{\hat{T},0}(\Delta)\left((d+\Delta\bar\Delta)^2G_\Delta(\sigma)+\sigma(1+2d+2\Delta\bar\Delta)G_\Delta'(\sigma)+\sigma^2G''_\Delta(\sigma)\right)\Big)\nonumber+\ldots\\
    T_3(\sigma)&=-\frac{2\pi}{\sigma}\int_{\frac{d}{2}+i\mathbb{R}}\frac{d\Delta}{2\pi i}\left(\frac{\varrho^{\mathcal{P}}_{\hat{T},2}(\Delta)}{4}\left(2\mathcal{G}_0(\sigma)+\sigma\mathcal{G}_1(\sigma)\right)+\varrho^{\mathcal{P}}_{\hat{T},0}(\Delta)\left(G_{\Delta}''(\sigma)+\sigma G_\Delta'''(\sigma)\right)\right)+\ldots\nonumber\\
    T_2(\sigma)&=2\pi\int_{\frac{d}{2}+i\mathbb{R}}\frac{d\Delta}{2\pi i}\Big(\frac{\varrho^{\mathcal{P}}_{\hat{T},2}(\Delta)}{(d+1)}\left((\sigma^2-1)\mathcal{G}_2(\sigma)+\sigma\mathcal{G}_1(\sigma)+\mathcal{G}_0(\sigma)\right)\\
    &\qquad\qquad\qquad\qquad\qquad\qquad-\varrho^{\mathcal{P}}_{\hat{T},0}(\Delta)\left((d+2+\Delta\bar\Delta)G''_\Delta(\sigma)+\sigma G'''_\Delta(\sigma)\right)\Big)+\ldots\nonumber\\
    T_1(\sigma)&=\frac{2\pi}{\sigma^2}\int_{\frac{d}{2}+i\mathbb{R}}\frac{d\Delta}{2\pi i}\Big(\varrho^{\mathcal{P}}_{\hat{T},2}(\Delta)\left(\mathcal{G}_0(\sigma)+\sigma\mathcal{G}_1(\sigma)+\sigma^2\mathcal{G}_2(\sigma)\right)\nonumber\\
    &\qquad\qquad\qquad\qquad+\varrho^{\mathcal{P}}_{\hat{T},0}(\Delta)\left(2G''_\Delta(\sigma)+4\sigma G'''_\Delta(\sigma)+\sigma^2 G''''_\Delta(\sigma)\right)\Big)+\ldots\nonumber
\end{align}
where the functions $\mathcal{G}_m$ are defined in (\ref{eq:Gdelta2exp}), primes are derivatives with respect to $\sigma$, and the dots stand for contributions from complementary and exceptional series. In particular, the complementary series contributions are exactly the same, with the only difference being the domain of integration. We checked these equations in the case of the free massive boson.

In two dimensions, we can find the analogous relations for the $\mathcal{T}_i$ functions by again extracting the coefficients of the tensor structures (\ref{eq:uniquestructs}) from (\ref{eq:stresstensorspectral2d}) and then using (\ref{eq:TtotildeT}). We obtain
\begin{align}
    \mathcal{T}_1(\sigma)=&(1-\sigma^2)\int_{\frac{1}{2}+i\mathbb{R}}\frac{d\Delta}{2\pi i}\frac{\varrho_\Theta^{\mathcal{P}}(\Delta)}{N(\Delta)}\Big((\Delta\bar\Delta+5)G''_\Delta(\sigma)+\sigma(5G'''_\Delta(\sigma)+\sigma G''''_\Delta(\sigma))\Big)\nonumber\\
    &+\frac{3}{4\pi}\varrho^{\mathcal{D}_2}_{\hat T}\frac{1+\sigma}{(1-\sigma)^3}+\text{complementary}\\
    \mathcal{T}_2(\sigma)=&(1-\sigma^2)\int_{\frac{1}{2}+i\mathbb{R}}\frac{d\Delta}{2\pi i}\frac{\varrho_\Theta^{\mathcal{P}}(\Delta)}{N(\Delta)}\Big((\Delta\bar\Delta+5)G_\Delta''(\sigma)+3\sigma G'''_\Delta(\sigma)\Big)+\frac{3}{4\pi}\varrho_{\hat T}^{\mathcal{D}_2}\frac{1+\sigma}{(1-\sigma)^2}\nonumber\\
    &+\text{complementary}\\
    \mathcal{T}_3(\sigma)=&\int_{\frac{1}{2}+i\mathbb{R}}\frac{d\Delta}{2\pi i}\frac{\varrho_\Theta^{\mathcal{P}}(\Delta)}{N(\Delta)}\Big((\Delta\bar\Delta+1)^2G_\Delta(\sigma)+(3+2\Delta\bar\Delta)\sigma G'_\Delta(\sigma)+(2-\sigma^2)G''_\Delta(\sigma)\nonumber\\
    &\qquad\qquad\qquad+2\sigma(1-\sigma^2)G_\Delta'''(\sigma)\Big)+\frac{3}{8\pi}\varrho^{\mathcal{D}_2}_{\hat T}\frac{1+2\sigma}{(1-\sigma)^2}+\text{complementary}\\
    \mathcal{T}_4(\sigma)=&\int_{\frac{1}{2}+i\mathbb{R}}\frac{d\Delta}{2\pi i}\frac{\varrho_\Theta^{\mathcal{P}}(\Delta)}{N(\Delta)}\Big((\Delta\bar\Delta+1)^2G_\Delta(\sigma)+(3+2\Delta\bar\Delta)\sigma G'_\Delta(\sigma)+(2+\sigma^2)G''_\Delta(\sigma)\Big)\nonumber\\
    &+\frac{3}{8\pi}\varrho_{\hat T}^{\mathcal{D}_2}\frac{1}{(1-\sigma)^2}+\text{complementary}\,,
    \label{eq:TTTTspectrals}
\end{align}
where here
\begin{equation}
    N(\Delta)\equiv\frac{(\Delta+1)^2(\bar\Delta+1)^2}{\pi}\,.
\end{equation}
By plugging in $\sigma=-1$ and using that the $n$-th derivative of the scalar propagator at antipodal separation is 
\begin{equation}
    \partial_\sigma^n G_\Delta(\sigma)\Big|_{\sigma=-1}=\frac{\Gamma(\Delta+n)\Gamma(\bar\Delta+n)}{2^{2+n}\pi n!}\,,
\end{equation}
we find that all the integrands for $\mathcal{T}_i(-1)$ go like $e^{i\pi\Delta}\varrho_\Theta^{\mathcal{P}}(\Delta)$ as $\Delta\to\frac{1}{2}+i\infty$. They thus converge if $\varrho_\Theta^{\mathcal{P}}(\Delta)$ does not grow exponentially in that same limit. This limit corresponds to the flat space limit, and in flat space spectral densities can only grow polynomially. We thus proved that the $\mathcal{T}_i$ functions are analytic around $\sigma=-1$. Moreover
\begin{equation}
\mathcal{T}_1(-1)=\mathcal{T}_2(-1)=0\,.
\end{equation}
\section{Details on the free scalar and the free fermion}
Here we show some computational details and checks of our formulas in the cases of a free massive scalar and a free massive Majorana fermion.
\subsection{Free massive scalar}
\label{subsec:detailsboson}
Consider the theory of a free massive scalar with $m^2R^2=\Delta_\phi(d-\Delta_\phi)$. We work in $d+1$ dimensions, but we are ultimately interested in taking the limit $d\to1$. We will thus ignore improvement terms in the stress tensor which arise from the conformal coupling in the action $\frac{d-1}{4d}\text{R}\phi^2$,\footnote{In this expression R is the Ricci scalar.} since in $d=1$ the coupling is zero.
\begin{equation}
    S=-\frac{1}{2}\int d^{d+1}x\sqrt{|g|}\left(g^{\mu\nu}\partial_\mu\phi\partial_\nu\phi+m^2\phi^2\right)\,.
\end{equation}
As written in the main text, the stress tensor of this theory, which we here report uplifted to embedding space, is
\begin{equation}
    T_{AB}=\nabla_A\phi\nabla_B\phi-\frac{1}{2}G_{AB}\left[\nabla^C\phi\nabla_C\phi+m^2\phi^2\right]\,.
\end{equation}
We can split it into its traceless part and its trace. For convenience, we introduce some auxiliary vectors $W$ which are null $(W^2=0)$ and tangent to the hypersurface in embedding space $(W\cdot Y=0)$, for the purpose of contracting all the indices while enforcing symmetricity and tracelessness \cite{Schaub:2023scu,Pethybridge:2021rwf,Loparco:2023rug,Sleight_2020,Sleight_2021,Sleight_20212,Costa:2014kfa}. Embedding space tensors are then traded for polynomials of $W$
\begin{equation}
    \hat T(W)\equiv W^AW^BT_{AB}=(W\cdot\nabla)\phi(W\cdot\nabla)\phi\,.
\end{equation}
To retrieve the expression of the traceless part of the stress tensor with indices, we act with the Todarov operator \cite{Schaub:2023scu,Pethybridge:2021rwf,Loparco:2023rug}
\begin{equation}
\begin{aligned}
    K_A\equiv&\frac{d-1}{2}\left[\partial_{W^A}-Y_A(Y\cdot\partial_{W})\right]+(W\cdot\partial_W)\partial_{W^A}-Y_A(Y\cdot\partial_W)(W\cdot\partial_W)\\
	&-\frac{1}{2}W_A\left[(\partial_W\cdot\partial_W)-(Y\cdot\partial_W)^2\right]\,,
\end{aligned}
\end{equation}
in the following way
\begin{equation}
    \hat T_{AB}=\frac{K_AK_B}{2(\frac{d-1}{2})_2}\hat T(W)\,.
\end{equation}
Finally, let us mention that the covariant derivative has to be modified to accomodate the use of the $W$ vectors
\begin{equation}
	\nabla_A=\partial_{Y^A}-Y_A(Y\cdot\partial_Y)-W_A(Y\cdot\partial_W)\,.
\end{equation}
The first thing we do is to check whether there is some range of parameters for which the principal series is the only contribution to the K\"allén-Lehmann decomposition of $\hat T$, such that we can apply the inversion formulae from \cite{Loparco:2023rug}. The criterion, also outlined in \cite{Loparco:2023rug}, is based on the fall-off of the components of the two-point function of $\hat T$ as we take $\sigma\to-\infty$. Let us write a generic two-point function of a spin 2 operator in index free formalism as
\begin{equation}
    \langle\mathcal{O}(Y_1,W_1)\mathcal{O}(Y_2,W_2)\rangle=\sum_{m=0}^2(W_1\cdot W_2)^{2-m}[(Y_1\cdot W_2)(Y_2\cdot W_1)]^{m}\mathcal{G}_m(\sigma)\,.
    \label{eq:tracelessbasis}
\end{equation}
Then, the criterion for the principal series being the only class of UIRs appearing in the spectral decomposition of this two-point function is that the fall-offs of the $\mathcal{G}_m$ functions respect the following inequality
\begin{equation}
    \lim_{\sigma\to-\infty}\mathcal{G}_{m}(\sigma)\sim|\sigma|^{-\omega_{m}-m}\,, \qquad \underset{m}{\text{min}}[\text{Re}(\omega_{m})]>\frac{d}{2}+2\,,
\end{equation}
When this condition is satisfied, the two-point function is square integrable when continued to EAdS, which ensures that harmonic functions in the principal series furnish a complete basis \cite{Camporesi:1994ga,Costa:2014kfa}, see section 4.3 in \cite{Loparco:2023rug} for a detailed discussion. The two-point function of interest to us is
\begin{equation}
    \langle\hat T(Y_1,W_1)\hat T(Y_2,W_2)\rangle=2\left[(W_1\cdot\nabla_1)(W_2\cdot\nabla_2)\langle\phi(Y_1)\phi(Y_2)\rangle\right]^2\,.
\end{equation}
The fall-offs of its components in the basis (\ref{eq:tracelessbasis}) are
\begin{equation}
    \text{min}\ \omega_0=\text{min}\ \omega_1=\text{min}\ \omega_2=2+2\text{min}(\text{Re}\Delta_\phi,\text{Re}\bar\Delta_\phi)\,.
\end{equation}
We can thus say that for $\text{min}(\text{Re}\Delta_\phi,\text{Re}\bar\Delta_\phi)>\frac{d}{4}$, the principal series is the only contribution to the spectral decomposition of $\hat T$. Let us start by assuming we are in this regime, which is satisfied when the free boson is in the principal series or in a portion of the complementary series $\Delta_\phi\in(\frac{d}{4},\frac{3d}{4})$. Then, we can decompose the traceless part of the stress tensor in the principal series only
\begin{equation}
\begin{aligned}
    \langle\hat T(Y_1,W_1)\hat T(Y_2,W_2)\rangle=2\pi\int_{\frac{d}{2}+i\mathbb{R}}\frac{d\Delta}{2\pi i}\Big[&\varrho_{\hat T,2}^{\mathcal{P}}(\Delta)G_{\Delta,2}(Y_1,Y_2;W_1,W_2)\\
    &+\varrho_{\hat T,0}^{\mathcal{P}}(\Delta)(W_1\cdot\nabla_1)^2(W_2\cdot\nabla_2)^2G_\Delta(\sigma)\Big]\,,
    \label{eq:mah}
\end{aligned}
\end{equation}
where we used the facts proven in Section \ref{subsec:spectralrelations} to exclude spin 1 contributions, and the explicit expression of $G_{\Delta,2}$ is given in (\ref{eq:Gdelta2exp}). Applying the inversion formulae from \cite{Loparco:2023rug}, specifically with the methods outlined in appendix H there, we compute the spectral densities in (\ref{eq:mah})
\begin{align}
    \varrho_{\hat T,2}^{\mathcal{P}}(\Delta)&=\frac{\lambda\sinh(\pi\lambda)\Gamma\left(\frac{2+\frac{d}{2}\pm i\lambda}{2}\right)^2}{2\pi^{3+\frac{d}{2}}\Gamma(\frac{d}{2}+2)\Gamma(2+\frac{d}{2}\pm i\lambda)}\prod_{\pm,\pm}\Gamma\left(\frac{2+\frac{d}{2}\pm i\lambda\pm 2i\lambda_\phi}{2}\right)\,,\label{eq:densitiesvarrho}\\
    \varrho^{\mathcal{P}}_{\hat T,0}(\Delta)&=\frac{\left((d-1)\Delta\bar\Delta+4\Delta_\phi\bar\Delta_\phi\right)^2\lambda\sinh(\pi\lambda)\Gamma\left(\frac{\frac{d}{2}\pm i\lambda}{2}\right)^2}{2^8\pi^{3+\frac{d}{2}}d^2(\Delta+1)^2(\bar\Delta+1)^2\Gamma(\frac{d}{2})\Gamma(\frac{d}{2}\pm i\lambda)}\prod_{\pm,\pm}\Gamma\left(\frac{\frac{d}{2}\pm i\lambda\pm 2i\lambda_\phi}{2}\right)\,,\nonumber
\end{align}
where we are using $\Delta=\frac{d}{2}+i\lambda$ and $\Delta_\phi=\frac{d}{2}+i\lambda_\phi$ and the radius has been set to 1. The integral in (\ref{eq:mah}) can then be checked numerically. We also independently compute $\varrho_{\Theta\hat T}$ and $\varrho_{\Theta}$ and we check that the identities (\ref{eq:spectralrelations}) are verified. Using those identities and more in general what is discussed in Section \ref{subsec:spectralrelations}, we can thus write the spectral decomposition of the full stress tensor two-point function for the free boson in the regime where $\text{min}(\text{Re}\Delta_\phi,\text{Re}\bar\Delta_\phi)>\frac{d}{4}$:
\begin{equation}
\begin{aligned}
    \langle T^{AB}(Y_1)T^{CD}(Y_2)\rangle=2\pi\int_{\frac{d}{2}+i\mathbb{R}}\frac{d\Delta}{2\pi i}\Big[&\varrho^{\mathcal{P}}_{\hat T,2}(\Delta)G^{AB,CD}_{\Delta,2}(Y_1,Y_2)\\
    &+\frac{\varrho^{\mathcal{P}}_\Theta(\Delta)}{d^2(\Delta+1)^2(\bar\Delta+1)^2}\Pi_1^{AB}\Pi_2^{CD}G_\Delta(\sigma)\Big]\,.\label{eq:TTdp1}
\end{aligned}
\end{equation}
Now we start the continuation to $d=1$. First of all, let us write the explicit form of $G_{\Delta,2}$, the free propagator of a massive traceless and transverse spin 2 field in de Sitter. In index-free notation, it is the solution to 
\begin{equation}
\left(\nabla^2_1-\Delta\bar\Delta-2\right)G_{\Delta,2}(Y_1,Y_2;W_1,W_2)=0\,, \qquad (K_1\cdot\nabla_1)G_{\Delta,2}(Y_1,Y_2;W_1,W_2)=0\,,
\end{equation}
with the extra condition of finiteness at antipodal separation. Because of $SO(1,d+1)$ invariance and the tangential condition $W_i\cdot Y_i=0$, we can express the solution in terms of three scalar functions multiplying the elements of a polynomial of dot products involving the $W$ vectors
\begin{equation}
   G_{\Delta,2}(Y_1,Y_2;W_1,W_2)=\sum_{m=0}^2(W_1\cdot W_2)^{2-m}[(W_1\cdot Y_2)(W_2\cdot Y_1)]^{m}\mathcal{G}_m(\sigma)\,,
\end{equation}
with \cite{Loparco:2023rug}
\begin{align}        \frac{\mathcal{G}_0(\sigma)}{N(\Delta)}&=8\left(2d(\mathbf{F}^{(0)}+\sigma\mathbf{F}^{(1)})+(\sigma^2d-1)\mathbf{F}^{(2)}\right)\label{eq:Gdelta2exp}\,,\\
        \frac{\mathcal{G}_1(\sigma)}{N(\Delta)}&=8\Big(2d(d+1)\mathbf{F}^{(1)}+\sigma d(5+3d+\Delta\bar\Delta)\mathbf{F}^{(2)}+(\sigma^2d-1)(\Delta+2)(\bar\Delta+2)\mathbf{F}^{(3)}\Big)\nonumber\,,\\
        \frac{\mathcal{G}_2(\sigma)}{N(\Delta)}&=4(d)_3\mathbf{F}^{(2)}+(\Delta+2)(\bar\Delta+2)(4d(d+2)\sigma\mathbf{F}^{(3)}+(\sigma^2d-1)(\Delta+3)(\bar\Delta+3)\mathbf{F}^{(4)})\,,\nonumber
    \end{align}
where we use a shorthand notation for some regularized hypergeometric functions
\begin{equation}
	\mathbf{F}^{(a)}\equiv\mathbf{F}\left(\Delta+a,\bar\Delta+a,\frac{d+1}{2}+a,\frac{1+\sigma}{2}\right)\,, 
\end{equation}
and here
\begin{equation}
	N(\Delta)\equiv\frac{(\Delta+1)(\bar\Delta+1)\Gamma(\Delta)\Gamma(\bar\Delta)}{2^{d+5}\pi^{\frac{d+1}{2}}d(\Delta-1)(\bar\Delta-1)}\,.
 \label{eq:Ndelta1}
\end{equation}
The index-open form of this propagator is then retrieved as
\begin{equation}
    G_{\Delta,2}^{AB,CD}(Y_1,Y_2)=\frac{K_1^AK_1^BK_2^CK_2^D}{4\left(\frac{d-1}{2}\right)_2^2}G_{\Delta,2}(Y_1,Y_2;W_1,W_2)\,.
\end{equation}
Notably, the normalization factor (\ref{eq:Ndelta1}) has simple poles at $\Delta=1$ and $\bar\Delta=1$, or equivalently at $\lambda=\pm i\frac{d-2}{2}$. When continuing in the number of dimensions, these poles will cross the integration contour over the principal series in (\ref{eq:TTdp1}) when passing by $d=2$. The residues on their positions need to be added by hand in order to retrieve the correct K\"allén-Lehmann representation in $d=1$. In \cite{Loparco:2023rug}, we showed that on these spurious poles, propagators and spectral densities associated to different spins are related to each other. The relations relevant here are
\begin{equation}
\begin{aligned}
	 \underset{\Delta=d-1}{\text{Res}}G_{\Delta,2}(Y_1,Y_2;W_1,W_2)&=\frac{2-d}{d}(W_1\cdot\nabla_1)^2(W_2\cdot\nabla_2)^2G_{d+1,0}(\sigma)\,,\\
	\varrho_{\hat T,2}(d-1)&=d (d-2)\underset{\Delta=d+1}{\text{Res}}\varrho_{\hat T,0}(\Delta)\,.
\end{aligned}
\end{equation}
Using the conservation relations (\ref{eq:spectralrelations}), we can further say 
\begin{equation}
	\varrho_{\hat T,2}(d-1)=\frac{d-2}{d(d+2)^3}\left((d+2)\partial_\Delta\varrho_\Theta(d+1)-2\varrho_\Theta(d+1)\right)\,.
\end{equation}
We thus see that in two dimensions ($d=1$) there will be the appearance of a UIR with $\Delta=2$ in the K\"allén-Lehmann representation of the traceless part of the stress tensor of a free massive boson. In particular, in this case $\varrho_\Theta(2)=0$, and what we are left with is
\begin{equation}
	\varrho^{\mathcal{D}_2}_{\hat T}=\frac{4\pi}{9}\partial_\Delta\varrho^{\mathcal{P}}_\Theta(2)=\frac{\lambda_\phi m^2}{3}\text{csch}(2\pi\lambda_\phi)\,,
\label{eq:spectraldiscrete}
\end{equation}
where $m^2=\frac{1}{4}+\lambda_\phi^2$ ($R=1$ here). Finally, in $d=1$, the following identities are true
\begin{equation}
\begin{aligned}
	\left(Y_1\cdot W_2^\pm\right)\left(Y_2\cdot W_1^\pm\right)&=(\sigma+1)\left(W_1^\pm\cdot W_2^\pm\right)\,,\\
	\left(Y_1\cdot W_2^\mp\right)\left(Y_2\cdot W_1^\pm\right)&=(\sigma-1)\left(W_1^\pm\cdot W_2^\mp\right)\,.
\label{eq:pmWid}
\end{aligned}
\end{equation}
where $\pm$ stands for the $SO(1,2)$ chirality. These identities stem from the fact that spin $J$ tensors have only two independent components in two dimensions, corresponding to two $SO(1,2)$-inequivalent $W^A$. Every two-point function of spin $J$ operators in two dimensions can be then decomposed in two components, one proportional to $(W_1^\pm\cdot W_2^\pm)^J$ and one proportional to $(W_1^\pm\cdot W_2^\mp)^J$. The second one is vanishing except if the theory violates parity.

It can be checked that, using (\ref{eq:pmWid}) in (\ref{eq:Gdelta2exp}), both components of the two-point function $G_{\Delta,2}$ vanish in two dimensions. All in all, the spectral decomposition of the stress tensor of a free massive boson in the principal series in two dimensions, obtained by continuing in $d$ from (\ref{eq:TTdp1}), is
\begin{equation}
\begin{aligned}
    \langle T^{AB}(Y_1)T^{CD}(Y_2)\rangle=&2\pi\int_{\frac{1}{2}+i\mathbb{R}}\frac{d\Delta}{2\pi i}\ \frac{\varrho^{\mathcal{P}}_\Theta(\Delta)}{(\Delta+1)^2(\bar\Delta+1)^2}\Pi_1^{AB}\Pi_2^{CD}G_{\Delta}(\sigma)\\
    &+\varrho^{\mathcal{D}_2}_{\hat T}\Pi_1^{AB}\Pi^{CD}_2G_{\Delta=2}(\sigma)\,,
\label{eq:spectralbosonTT2}
\end{aligned}
\end{equation}
with the spectral densities given by setting $d=1$ in (\ref{eq:densitiesvarrho}) and (\ref{eq:spectraldiscrete}). 
\paragraph{Complementary series contributions}
Until now we had assumed the free scalar sits in the range $\text{min}(\text{Re}\Delta_\phi,\text{Re}\bar\Delta_\phi)>\frac{1}{4}$, or equivalently $m^2>3/16$. We can analytically continue (\ref{eq:spectralbosonTT2}) beyond that regime. From the explicit expression of $\varrho_\Theta(\Delta)$ in (\ref{eq:densitiesvarrho}) we see that poles at $\lambda=\pm2\lambda_\phi+\frac{i}{2}$ cross the integration contour over the principal series when $|\text{Im}\lambda_\phi|>\frac{1}{4}$. Summing the residues on these poles, we obtain the full decomposition
\begin{equation}
\begin{aligned}
    \langle T^{AB}(Y_1)T^{CD}(Y_2)\rangle=&2\pi\int_{\frac{1}{2}+i\mathbb{R}}\frac{d\Delta}{2\pi i}\frac{\varrho^{\mathcal{P}}_\Theta(\Delta)}{(\Delta+1)^2(\bar\Delta+1)^2}\Pi_1^{AB}\Pi_2^{CD}G_{\Delta}(\sigma)\\
    &+\int_0^1d\Delta\frac{\varrho^{\mathcal{C}}_\Theta(\Delta)}{(\Delta+1)^2(\bar\Delta+1)^2}\Pi_1^{AB}\Pi_2^{CD}G_{\Delta}(\sigma)\\
    &+\varrho^{\mathcal{D}_2}_{\hat T}\Pi_1^{AB}\Pi^{CD}_2G_{\Delta=2}(\sigma)\,,
\label{eq:spectralTT}
\end{aligned}
\end{equation}
with
\begin{equation}
\begin{aligned}
    \varrho^{\mathcal{C}}_\Theta(\Delta)&=-\delta(\Delta-2\Delta_\phi+1)\theta\left(\Delta_\phi-\frac{3}{4}\right)\frac{(\Delta+1)^2\bar\Delta\cos(\pi\Delta)\Gamma(\frac{3}{2}-\Delta)\Gamma(\frac{3-\Delta}{2})\Gamma(\frac{\Delta}{2})^2}{2^{4-\Delta}\pi^2R^4\Gamma(1-\frac{\Delta}{2})}\,,
    \label{eq:rhocrhodboson2}
\end{aligned}
\end{equation}
and where $\theta(x)$ is a Heaviside theta function.

Notice that all these extra terms can be added as a modification of the original contour of integration, becoming
\begin{equation}
\langle T^{AB}(Y_1)T^{CD}(Y_2)\rangle=2\pi\int_{\gamma}\frac{d\Delta}{2\pi i}\ \frac{\varrho^{\mathcal{P}}_\Theta(\Delta)}{(\Delta+1)^2(\bar\Delta+1)^2}\Pi_1^{AB}\Pi_2^{CD}G_{\Delta}(\sigma)\,,
\end{equation}
where $\gamma$ is the contour represented in blue in figure \ref{fig:contours}.
\paragraph{The c-functions}
By using the techniques outlined in Appendix \ref{subsec:Tifuncs}, we compute the function $c_1(R)$ from its definition (\ref{eq:defcR}). For the free boson in two-dimensions we obtain, in particular
\begin{equation}
    \mathcal{T}_3(-1)=\frac{m^4}{128}\text{csc}^2(\pi\Delta_\phi)\,, \qquad \mathcal{T}_4(-1)=3\frac{m^4}{128}\text{csc}^2(\pi\Delta_\phi)\,.
\end{equation}
Using (\ref{eq:defcR}) we thus get $c_1(R)=0\,,$ which is due to the IR divergences associated to massless scalars in de Sitter, affecting the trace of the stress tensor. The second c-function is reported in the main text (\ref{eq:c2boson}).
\subsection{Free massive fermion}
\label{subsec:detailsfermion}
Consider the theory of a free massive Majorana fermion in two-dimensional de Sitter space, described by the action 
\begin{equation}
    S=-\frac{1}{2}\int d^2x\sqrt{|g|}\bar\Psi\left(\slashed{\nabla}+m\right)\Psi\,.
    \label{eq:actionfermion2}
\end{equation}
The only spin $\frac{1}{2}$ UIRs are in the principal series, with mass and conformal weight related through $\Delta=\frac{1}{2}+imR$, with $m>0$ \cite{Dirac:1935zz,Gursey:1963ir}. We choose to work with conventions in which $\Psi$ is a real bispinor
\begin{equation}
    \Psi=\begin{pmatrix}
        \psi_1 \\ \psi_2
    \end{pmatrix}\,,
\end{equation}
where $\psi_1$ and $\psi_2$ are real Grassmann functions.
Moreover, it is useful to go to local coordinates, and we choose the flat slicing metric $ds^2=R^2\frac{-\mathrm{d}\eta^2+\mathrm{d}y^2}{\eta^2}$. Then, we choose the (flat) gamma matrices to be
\begin{equation}
    \gamma_0=\begin{pmatrix}
        0 & 1 \\ -1 & 0
    \end{pmatrix}\,, \qquad \gamma_1=\begin{pmatrix}
        0 & 1 \\ 1 & 0
    \end{pmatrix}\,.
\end{equation}
The corresponding gamma matrices in de Sitter are given by $\Gamma^\mu=e^\mu_a \gamma^a$, with the zweibein satisfying $e^a_\mu e^b_\nu \eta_{ab}=g_{\mu\nu}$. With these conventions, the charge conjugation matrix, defined by
\begin{equation}
    C\gamma_\mu C^{-1}=-\gamma_\mu^T\,,
\end{equation}
can be chosen to be $C=\begin{pmatrix}
    0 & 1 \\ -1 & 0
\end{pmatrix}\,.$ Then, we have that $
    \bar\Psi=\begin{pmatrix}
        -\psi_2 & \psi_1
    \end{pmatrix}\,,$ and the two-point function is
\begin{equation}
   \langle\Psi(x_1)\bar\Psi(x_2)\rangle=\begin{pmatrix}
       -\langle\psi_1(x_1)\psi_2(x_2)\rangle & \langle\psi_1(x_1)\psi_1(x_2)\rangle\\
       -\langle\psi_2(x_1)\psi_2(x_2)\rangle & \langle\psi_2(x_1)\psi_1(x_2)\rangle
   \end{pmatrix}\,.
   \label{eq:free2ptfermion}
\end{equation}
As explained in the main text, the trace of the stress tensor in this theory is
\begin{equation}
    \Theta(x)=-\frac{m}{2}\bar\Psi\Psi(x)=-m\psi_1\psi_2(x)\,,
\end{equation}
and the associated two-point function is
\begin{equation}
    \langle\Theta(x_1)\Theta(x_2)\rangle=m^2(\langle\psi_1(x_1)\psi_2(x_2)\rangle\langle\psi_2(x_1)\psi_1(x_2)\rangle-\langle\psi_1(x_1)\psi_1(x_2)\rangle\langle\psi_2(x_1)\psi_2(x_2)\rangle)\,.
\end{equation}
The entries of the matrix (\ref{eq:free2ptfermion}) that solve the equations of motion \cite{Henningson:1998cd,Schaub:2023scu}
\begin{equation}
    \left(\slashed{\nabla}+m\right)\Psi=0 \qquad \longrightarrow \qquad \left(\eta\gamma^\mu\partial_\mu+\frac{1}{2}\gamma_0+m\right)\Psi=0\,,
\end{equation}
were given in eq. (\ref{eq:freefermion}) \cite{Schaub:2023scu,Pethybridge:2021rwf}.

We are now going to show that, in the flat space limit, we reproduce the correct two-point function, thus providing an independent check of the normalization presented in the references \cite{Schaub:2023scu,Pethybridge:2021rwf}. 
\paragraph{Flat space limit}
Let us focus on
\begin{equation}
    \begin{aligned}
    \mathcal{G}^-(\sigma)\equiv-\langle\psi_1(x_1)\psi_2(x_2)\rangle &=\frac{i[(\eta_1+\eta_2)+(y_1-y_2)]}{\sqrt{\eta_1\eta_2}}G^-_m(\sigma)\,,  \\
    \mathcal{G}^+(\sigma)\equiv\langle\psi_1(x_1)\psi_1(x_2)\rangle&=\frac{[(\eta_1-\eta_2)+(y_1-y_2)]}{\sqrt{\eta_1\eta_2}}G^+_m(\sigma)\,,
    \end{aligned}
\end{equation}
with $G^+_m$ and $G^-_m$ given in (\ref{eq:gplusgminus}).

As usual, we start by taking $\eta\to t-R$ and $y\to x$. Then
\begin{equation}
    \sigma=\frac{\eta_1^2+\eta_2^2-(y_1-y_2)^2}{2\eta_1\eta_2}\to 1-\frac{-(t_1-t_2)^2+(x_1-x_2)^2}{2R^2}\equiv1-\frac{x^2}{2R^2}\,.
\end{equation}
After some simplifications, we obtain
\begin{equation}
    \begin{aligned}
        \mathcal{G}^-(\sigma)&\to-\frac{m^2R}{4}\ \text{csch}(\pi mR)\ _2F_1\left(1-imR,1+imR,2,1-\frac{x^2}{4R^2}\right)\,,\\
    \mathcal{G}^+(\sigma)&\to\frac{m\text{csch}(\pi mR)}{8R}(t_1-t_2+x_1-x_2)\ _2F_1\left(1-imR,1+imR,1,1-\frac{x^2}{4R^2}\right)\,.
    \end{aligned}
\end{equation}
We use the following Barnes representation of the regularized hypergeometric function
\begin{equation}
\begin{aligned}
    &\mathbf{F}(a,b,c,z)=\frac{\int_{\mathbb{R}+i\epsilon} ds\Gamma(a+is)\Gamma(b+is)\Gamma(c-a-b-is)\Gamma(-is)(1-z)^{is}}{2\pi\Gamma(a)\Gamma(b)\Gamma(c-a)\Gamma(c-b)}\,.
    \label{eq:barnes}
\end{aligned}
\end{equation}
We can apply it directly to $\mathcal{G}^-$ without any issues.  Using $\Gamma(a\pm ib)\equiv \Gamma(a+ib)\Gamma(a-ib)$, we write
\begin{equation}
    \begin{aligned}
        \mathcal{G}^-(\sigma)\to-\frac{m^2R\ \text{csch}(\pi mR)}{8\pi\Gamma(1\pm imR)^2}\int_{\mathbb{R}+i\epsilon}ds\ \Gamma(1\pm imR+is)\Gamma(-is)^2\left(\frac{x_{12}^2}{4R^2}\right)^{is}
    \end{aligned}
\end{equation}
For $\mathcal{G}^+$, instead, there is a subtlety: the contour in (\ref{eq:barnes}) does not actually separate the two series of poles in the gamma functions. We thus need to introduce a regulator which we take to be $\alpha\geq1$ and we eventually will take to 0, and write
\begin{equation}
    \mathcal{G}_{(\alpha)}^+(\sigma)\equiv\frac{m\text{csch}(\pi mR)}{8R}(t_1-t_2+x_1-x_2)\mathbf{F}\left(1-imR,1+imR,1+\alpha,1-\frac{x^2}{4R^2}\right)\,.
\end{equation}
Then, the Barnes representation for the regulated $\mathcal{G}_{(\alpha)}^+(\sigma)$ reads
\begin{equation}
\begin{aligned}
    \mathcal{G}_{(\alpha)}^+(\sigma)\to&\frac{m\text{csch}(\pi mR)(t_1-t_2+x_1-x_2)}{16\pi R\Gamma(\pm imR)\Gamma(1\pm imR)}\\
    &\times\int_{\mathbb{R}+i\epsilon}ds\ \Gamma(1\pm imR+is)\Gamma(-is)\Gamma(\alpha-1-is)\left(\frac{x^2}{4R^2}\right)^{is}
\end{aligned}
\label{eq:galpharep}
\end{equation}
where we already took the regulator to zero where it didn't cause problems. 
Now we take the large radius limit. In this limit, 
\begin{equation}
\Gamma(a\pm ibR)\to 2\pi e^{-\pi bR}(bR)^{2a-1}\,,
\end{equation}
and only the growing part of csch$(\pi m R)$ matters
\begin{equation}
    \begin{aligned}
        \mathcal{G}^-(\sigma)&\to-\frac{m}{8\pi^2}\int_{\mathbb{R}+i\epsilon}ds\left(\frac{m^2x^2}{4}\right)^{is}\Gamma(-is)^2\,,\\
    \mathcal{G}_{(\alpha)}^+(\sigma)&\to\frac{m^2}{16\pi^2}((t_1-t_2)+(x_1-x_2))\int_{\mathbb{R}+i\epsilon}ds\left(\frac{m^2x^2}{4}\right)^{is}\Gamma(\alpha-1-is)\Gamma(-is)\,.
    \end{aligned}
\end{equation}
Here we recognize the Barnes representations of the modified Bessel function of the second kind
\begin{equation}
\begin{aligned}
    K_\nu(z)=\frac{1}{4\pi i}\left(\frac{z}{2}\right)^\nu\int_{c+i\mathbb{R}}dt\ \Gamma(t)\Gamma(t-\nu)\left(\frac{z}{2}\right)^{-2t}\,,
\label{eq:Knuinteg}
\end{aligned}
\end{equation}
with $c>\text{max}(\text{Re}(\nu),0).$ For $\mathcal{G}_-$, the result is spot on. For $\mathcal{G}_+^{(\alpha)}$, the validity of the integral representation (\ref{eq:Knuinteg}) depends on $\alpha$, specifically $\alpha\geq1$, which is precisely the values for which (\ref{eq:galpharep}) is valid. We can thus substitute also here the Bessel function, and we obtain
\begin{equation}
    \begin{aligned}
        \mathcal{G}^-(\sigma)&\to -\frac{m}{2\pi}K_0(m|x|)\,,\\
    \mathcal{G}_{(\alpha)}^+(\sigma)&\to \frac{m^2}{4\pi}\left(\frac{2}{m|x|}\right)^{1-\alpha}(t_1-t_2+x_1-x_2)K_{1-\alpha}(m|x|)\,.
    \end{aligned}
\end{equation}
The Bessel function is analytic in its order. We can thus now continue to $\alpha=0$ and obtain
\begin{equation}
    \begin{aligned}
        \mathcal{G}^-(\sigma)&\to -\frac{m}{2\pi}K_0(m|x|)\,,\\
    \mathcal{G}^+(\sigma)&\to \frac{m}{2\pi}\frac{(t_1-t_2+x_1-x_2)}{|x|}K_{1}(m|x|)\,.
    \end{aligned}
\end{equation}
Summarizing, we have shown that in the flat space limit
\begin{equation}
    \langle\Psi(x_1)\bar\Psi(x_2)\rangle\to\frac{m}{2\pi}\begin{pmatrix}
        -K_0(m|x|) & \frac{t_1+x_1-t_2-x_2}{|x|}K_1(m|x|)\\
        \frac{t_1-x_1-t_2+x_2}{|x|}K_1(m|x|) & -K_0(m|x|)
    \end{pmatrix}\,,
\end{equation}
precisely matching the canonical normalization (see for example \cite{Bander:1976qe}).
\bibliography{paper}
\bibliographystyle{utphys}
\end{document}